\documentclass[aps,physrev,preprint,superscriptaddress,nofootinbib]{revtex4-2}

\usepackage[
    colorlinks=true,
    citecolor=blue,
    linkcolor=blue,
    urlcolor=magenta
]{hyperref}

\usepackage[T1]{fontenc}
\usepackage{diagbox}
\usepackage{adjustbox}
\usepackage{multirow}
\usepackage{pifont}
\usepackage{array}
\usepackage{mathtools}
\usepackage{amsfonts}
\usepackage{amssymb}
\usepackage[dvipsnames]{xcolor}
\usepackage{tcolorbox}
\usepackage{float}
\usepackage{placeins}
\usepackage{braket}
\usepackage{slashed}
\usepackage{makecell}
\usepackage{ytableau}
\usepackage{bbold}
\usepackage{tabularx}
\usepackage{bm}
\usepackage{longtable}
\usepackage{graphicx}
\usepackage{booktabs}

\newcolumntype{L}[1]{>{\raggedright\arraybackslash}p{#1}}
\newcolumntype{R}[1]{>{\raggedleft\arraybackslash}p{#1}}

\newcommand{\classcount}[2]{%
  \(\begin{array}{@{}c@{}}
  #1\\[-0.5mm]
  \scriptstyle #2
  \end{array}\)
}

\newcommand{\vev}{v_\text{EW}}
\newcommand{\vh}{v_\mathcal{H}}
\newcommand{\vXi}{v_\Xi}

\usepackage[normalem]{ulem}
\newcommand{\stkout}[1]{\ifmmode\text{\sout{\ensuremath{#1}}}\else\sout{#1}\fi}

\newcommand{\beq}{\begin{equation}}
\newcommand{\eeq}{\end{equation}}
\newcommand{\bea}{\begin{eqnarray}}
\newcommand{\eea}{\end{eqnarray}}

\begin{document}

\title{Lepton-Number Violation in Higgs Effective Field Theory: Complete Operator Basis and Type-II Seesaw Matching}

\author{Zizhou Ge}
\email{gezizhou@snnu.edu.cn}
\affiliation{School of Physics and Information Technology, Shaanxi Normal University,
Xi'an 710119, People’s Republic of China}

\author{Huayang Song}
\email{huayangs1990@ibs.re.kr}
\affiliation{Particle Theory and Cosmology Group, Center for Theoretical Physics of the Universe, Institute for Basic Science (IBS), Daejeon, 34126, Republic of Korea}

\author{Xia Wan}
\email{wanxia@snnu.edu.cn}
\thanks{The authors are listed in alphabetical order by last name.}
\affiliation{School of Physics and Information Technology, Shaanxi Normal University,
Xi'an 710119, People’s Republic of China}

\date{\today}

\begin{abstract}
We construct a complete and nonredundant basis of lepton-number-violating (LNV) operators in the Higgs effective field theory (HEFT) through $\mathcal{O}(p^4)$, including their full flavor multiplicities. Using a complex dressed spurion that encodes the $B-L$ charge, we classify the $\Delta L=2$ and $\Delta L=4$ sectors, including baryon-number-violating operators, and independently validate the counting using Hilbert-series methods. As a concrete application, we present the first complete tree-level matching of the type-II seesaw model onto HEFT through $\mathcal{O}(p^4)$, covering the bosonic, lepton-number-conserving, and LNV sectors. Using the nonlinear field representation and primary-HEFT power counting, we retain the full dependence of the matching coefficients on the independent scalar masses, the neutral-scalar mixing, and the doublet and triplet vacuum expectation values. We compare our results with the corresponding real-triplet matching and a broken-phase effective field theory. We further discuss implications for LNV processes, Higgs physics, and electroweak observables.
\end{abstract}

\maketitle

\tableofcontents
\section{Introduction}

Lepton-number violation (LNV) is an important feature of many extensions of the Standard Model (SM), particularly those in which neutrinos acquire Majorana masses. At energies below the masses of the new particles, their effects can be described within effective field theory (EFT)~\cite{Weinberg:1978kz,Georgi:1993mps}. In the Standard Model Effective Field Theory (SMEFT)~\cite{Henning:2014wua,Brivio:2017vri}, the leading $\Delta L=2$ interaction is described by the dimension-five Weinberg operator, while additional LNV interactions appear at higher mass dimensions~\cite{Lehman:2014jma,Liao:2016hru,Li:2020xlh}. The classification of these operators, including their flavor symmetries and redundancies, has provided a useful basis for studying LNV processes such as neutrinoless double-beta decay, as well as for EFT matching and renormalization-group analyses.

The SMEFT description assumes a linear realization of electroweak symmetry, in which the physical Higgs boson and the electroweak Goldstone bosons are embedded in an $SU(2)_L$ doublet. This structure imposes relations among interactions involving different numbers of Higgs and Goldstone fields at a given order in the SMEFT expansion. Although SMEFT provides an economical description of decoupling new physics, its expansion may become inefficient when nondecoupling effects associated with electroweak symmetry breaking are important. In such cases, the Higgs Effective Field Theory (HEFT)~\cite{Feruglio:1992wf,Grinstein:2007iv,Alonso:2012px,Buchalla:2013rka,Alonso:2025ksv} provides a more general description, in which the electroweak Goldstone bosons are realized nonlinearly and the physical Higgs boson is treated as an electroweak singlet. The interactions are organized according to the chiral expansion, while undifferentiated Higgs fields enter through general coefficient functions of $h/v_{\rm EW}$. In particular, the Majorana neutrino mass term and its associated Higgs interactions in HEFT are not restricted by the relations inherited from the dimension-five Weinberg operator in SMEFT. This additional freedom also extends to higher-order LNV interactions.

Complete and nonredundant HEFT operator bases have been constructed in the lepton-number-conserving (LNC) sector~\cite{Sun:2022ssa}, while Hilbert-series methods have been employed to count operators that violate baryon and/or lepton number~\cite{Graf:2022rco}. However, to the best of our knowledge, an explicit, complete, and nonredundant basis of LNV HEFT operators through $\mathcal{O}(p^4)$, including their full flavor structures, has not yet been presented. Constructing such a basis would extend the HEFT operator classification to the LNV sector and provide a foundation for comparing the corresponding interactions with those in SMEFT, following previous comparisons of operator structures~\cite{Brivio:2013pma,Gavela:2014vra,Brivio:2016fzo,Merlo:2016prs} and EFT descriptions derived from explicit UV models~\cite{Dawson:2023ebe,Song:2025kjp,Ge:2026qfa}. It would also facilitate the study of LNV processes such as neutrinoless double-beta decay and same-sign dilepton production.

As a concrete ultraviolet (UV) realization of LNV, we consider the type-II seesaw model~\cite{Konetschny:1977bn,Magg:1980ut,Cheng:1980qt,Mohapatra:1980yp}. Unlike the type-I and type-III seesaw models, whose heavy mediators are fermions, the type-II seesaw model introduces a complex $SU(2)_L$ scalar triplet $\Xi$ with hypercharge $Y=1$, adopting the convention $Q=T_3+Y$. The triplet couples to two lepton doublets through a Yukawa interaction and to the Higgs doublet through a trilinear interaction in the scalar potential. Assigning lepton number $L(\Xi)=-2$, the trilinear interaction violates lepton number by two units and induces a triplet vacuum expectation value (VEV), thereby generating Majorana neutrino masses through the triplet Yukawa interaction. Parametrically, the neutrino mass matrix satisfies $(m_\nu)_{pr}\propto y_\Delta^{pr}v_\Xi$, with $v_\Xi\sim\mu \vh^2/M_\Xi^2$ in the heavy-triplet regime, up to convention-dependent numerical factors, where $\vh$ is the doublet VEV and $M_\Xi$ denotes the triplet mass scale. Meanwhile, the triplet participates directly in electroweak gauge and scalar interactions and introduces additional neutral, singly charged, and doubly charged scalar states, whose spectrum and interactions have been studied extensively~\cite{Arhrib:2011uy}. Their characteristic signatures have been investigated at hadron colliders~\cite{Mitra:2016wpr,Du:2018eaw,Ducu:2024xxf,Babu:2022ycv,Englert:2026eou}, at photon and muon colliders~\cite{Das:2023tna,Das:2024kyk}, and in related gauge-extended triplet scenarios~\cite{Das:2024yvt}. When these additional states are sufficiently heavy, their effects can be encoded in bosonic, LNC fermionic, and LNV effective operators. Experimental measurements can constrain the corresponding Wilson coefficients, while matching relates them to the masses and couplings of the triplet. The type-II seesaw model therefore provides a useful UV example for investigating LNV together with its connections to electroweak symmetry breaking and lepton-flavor physics.

The matching of the type-II seesaw model onto SMEFT has been extensively investigated. Earlier studies of scalar-triplet extensions were presented in Refs.~\cite{Dawson:2017vgm,Corbett:2017ieo,Dawson:2020oco}, while the complete one-loop matching through dimension six was performed in Refs.~\cite{Du:2022vso,Li:2022ipc}. These studies demonstrate how higher-dimensional effective operators encode information about the underlying UV dynamics beyond the Majorana neutrino mass term. Related one-loop matching results for the hybrid type-(I+II) seesaw model were obtained in Ref.~\cite{Zhang:2022osj}. At dimension eight, one-loop contributions from heavy scalars have been calculated for the complex $Y=1$ scalar triplet in a Green's basis, including both bosonic and fermionic operators and a discussion of their reduction to nonredundant bases~\cite{Adhikary:2025gdh}. More recently, a broken-phase EFT description was developed by integrating out the heavy physical states after electroweak symmetry breaking~\cite{Liao:2025bmn}. Despite these developments, a complete matching of the type-II seesaw model onto HEFT has not, to the best of our knowledge, been presented. Such a calculation provides a unified description of the bosonic, LNC fermionic, and LNV interactions generated by the type-II seesaw model within the nonlinear electroweak realization and allows comparisons with the existing SMEFT and broken-phase EFT descriptions.

In this work, we pursue two complementary objectives. First, we construct a complete and nonredundant basis of LNV HEFT operators through $\mathcal{O}(p^4)$, including their full flavor multiplicities. Inspired by the dimension-five Weinberg operator in SMEFT, we introduce a complex dressed spurion that encodes the $B-L$ charge and custodial orientation of the LNV insertion. We classify the $\Delta L=2$ and $\Delta L=4$ operators, together with the baryon-number-violating LNV structures, and independently verify the operator counting using Hilbert-series methods. Second, we perform, to the best of our knowledge, the first complete tree-level matching of the type-II seesaw model onto HEFT through $\mathcal{O}(p^4)$, including the bosonic, LNC fermionic, and LNV sectors. Employing the nonlinear field representation and the primary-HEFT power-counting scheme, we retain the dependence on the independent heavy-scalar masses, the neutral-scalar mixing angle, and the triplet VEV without introducing additional expansions in these parameters. We compare the results with the real-triplet extension and the broken-phase EFT, and discuss representative implications for LNV processes and Higgs and electroweak observables.

The remainder of this paper is organized as follows. In Sec.~\ref{sec:LNV}, we introduce the HEFT building blocks, construct the complete LNV operator basis, and verify the operator counting using Hilbert-series methods. In Sec.~\ref{sec:matching}, we present the type-II seesaw model, its complete tree-level matching onto HEFT through $\mathcal{O}(p^4)$, and a comparison with the broken-phase EFT. In Sec.~\ref{sec:pheno}, we discuss the phenomenological implications of the resulting effective operators. We summarize our findings in Sec.~\ref{sec:con}. The appendices contain details of the operator reduction, the HEFT-to-LEFT matching conditions relevant to neutrinoless double-beta decay, and the coefficient-by-coefficient comparison with the broken-phase EFT.

\section{Complete basis of lepton-number-violating HEFT operators\label{sec:LNV}}

\subsection{Nonlinear realization of electroweak symmetry and the LNC sector}
The SM Higgs sector possesses an approximate global $SU(2)_L\times SU(2)_R$ symmetry, which is spontaneously broken to the diagonal custodial subgroup $SU(2)_C$ by the Higgs VEV. This custodial symmetry is explicitly broken by the gauging of hypercharge and by the Yukawa interactions. Following the Callan--Coleman--Wess--Zumino (CCWZ) construction~\cite{Coleman:1969sm}, HEFT describes the electroweak Goldstone bosons as nonlinearly realized degrees of freedom, while the physical Higgs boson is treated as an electroweak singlet. The three Goldstone fields $\pi_i$ are collected into the unitary matrix $U\equiv \exp\left(\frac{i\pi_i\sigma_i}{v_{\rm EW}}\right)$, where $\sigma_i$ are the Pauli matrices. Under the global symmetry, $U$ transforms as $U\to g_L U g_R^\dagger$.

At leading order, the part of the LNC HEFT Lagrangian relevant to the present analysis is
\beq
\begin{aligned}[b]
\mathcal{L}_{\text{HEFT}}^{\text{LO}} \supset{}& \frac{1}{2}\partial_\mu h \partial^\mu h - \mathcal{V}( h ) - \frac{\vev^2}{4} \mathcal{F}( h ) \braket{V_\mu V^\mu } \\
& - \frac{\vev^2}{4}\mathcal{G}( h ) \braket{T V_\mu} \braket{T V^\mu} \\
& - \frac{\vev}{\sqrt{2}}\Bigl(\bar{Q}_L U\mathcal{Y}_Q( h ) Q_R + \bar{L}_L U\mathcal{Y}_L( h ) L_R + \text{h.c.}\Bigr),
\end{aligned}
\label{eq:HEFT}
\eeq
where $v_{\rm EW}=246~{\rm GeV}$ denotes the electroweak scale, $\langle\cdots\rangle$ denotes the trace over weak-isospin indices, and $D_\mu$ is the covariant derivative. We adopt the building blocks $V_\mu\equiv U D_\mu U^\dagger$ and $T\equiv U\sigma_3 U^\dagger$, where $T$ acts as a custodial-symmetry-breaking spurion. The
coefficient functions $\mathcal{V}(h)$, $\mathcal{F}(h)$, $\mathcal{G}(h)$, and $\mathcal{Y}_{Q/L}(h)$ are general polynomial functions of the Higgs field and may be expanded as
\begin{gather}
    \mathcal{V}( h )=\frac{1}{2}m_h^2 h^2\left[1+(1+\Delta\kappa_3)\frac{h}{\vev}+\frac{1}{4}(1+\Delta\kappa_4)\frac{h^2}{\vev^2}+\cdots\right], \label{kappa34def} \\
    \mathcal{F}( h )=1+2(1+\Delta a)\frac{h}{\vev}+(1+\Delta b)\frac{h^2}{\vev^2}+\cdots, \\
    \mathcal{G}( h )=\Delta\alpha+\Delta a^\slashed{C}\frac{h}{\vev}+\Delta b^\slashed{C}\frac{h^2}{\vev^2}+\cdots, \\
    \mathcal{Y}_Q( h )=\mathrm{diag}\biggl(\sum_n Y_U^{(n)}\frac{h^n}{\vev^n},\,\sum_n Y_D^{(n)}\frac{h^n}{\vev^n}\biggr),\ \mathcal{Y}_L( h )=\mathrm{diag}\biggl(0,\,\sum_n Y_\ell^{(n)}\frac{h^n}{\vev^n}\biggr).
\end{gather}

The first term in Eq.~\ref{eq:HEFT} is the Higgs kinetic term, while $\mathcal{V}(h)$ denotes the Higgs potential. The term proportional to $\mathcal{F}(h)$ contains the Goldstone kinetic term and, after electroweak symmetry breaking, generates the masses of the massive gauge bosons and their interactions with the Higgs field. The term proportional to $\mathcal{G}(h)$ explicitly breaks custodial symmetry and modifies the neutral gauge-boson sector. Finally, the Yukawa coefficient functions $\mathcal{Y}_Q(h)$ and $\mathcal{Y}_L(h)$ generate the fermion masses and the corresponding Higgs--fermion interactions. The vanishing upper entry of $\mathcal{Y}_L(h)$ reflects the absence of a right-handed neutrino in the HEFT considered here.

\subsection{The LNV spurion and operator basis}
However, the Lagrangian in Eq.~\eqref{eq:HEFT} contains only the leading-order LNC terms. Existing HEFT operator bases at higher chiral orders have also mainly focused on the LNC sector, while a complete and nonredundant basis of LNV operators remains to be constructed. To systematically incorporate LNV interactions into HEFT, we introduce a charged spurion inspired by the dimension-five Weinberg operator in SMEFT,
\begin{align}
    \mathcal{O}_{5}^{pr}=(L_p^T C\epsilon\mathcal{H})(\mathcal{H}^T\epsilon L_r).
\end{align}
where $\epsilon\equiv i\sigma_2$. Introducing the Higgs bidoublet $\Sigma=(\widetilde{\mathcal{H}},\mathcal{H})$, the Weinberg operator can be written as
\begin{align}
    L_{L p}^T C\epsilon\Sigma \sigma_-\epsilon\Sigma^T\epsilon L_{L r}.
\end{align}
where we define a projection operator $\sigma_-\equiv(\sigma_1-i\sigma_2)/2$, the lowering operator in $SU(2)_R$ space. The combination $\sigma_-\epsilon=(1-\sigma_3)/2$ projects onto the second column of $\Sigma$, namely $\mathcal{H}$, when it act from the right on $\Sigma$. So that
\begin{align}
    \Sigma \sigma_-\epsilon\Sigma^T=\mathcal{H}\mathcal{H}^T.
\end{align}

Using the polar decomposition $\Sigma=(v+h)U/\sqrt{2}$, we obtain
\begin{align}
    \mathcal{O}_{5}^{pr}=L_{L p}^T C\epsilon U \sigma_-\epsilon U^T\epsilon L_{L r}\mathcal{F}_W(h)=-L_{L p}^T C U^*\epsilon \sigma_- U^\dagger L_{L r}\mathcal{F}_W(h)
\end{align}
where $\mathcal{F}_W(h)\equiv(v+h)^2/2$ is the Higgs dependence of this illustrative SMEFT operator and we have used the $SU(2)$ pseudoreality identity $U^T\epsilon=\epsilon U^\dagger$ in the equality. In the general HEFT, the corresponding LNV operator may instead carry an independent coefficient function $\mathcal{F}_L^{pr}(h/v_{\rm EW})$ with arbitrary powers of the singlet Higgs field, unconstrained by this SMEFT example. Thus, in analogy with the usual dressed custodial spurion $U\sigma_3 U^\dagger$, the combination $U^*\epsilon \sigma_- U^\dagger$ can be interpreted as a dressed lepton-number-violating spurion in left space and selects a definite direction in right-handed custodial space.

To encode both this custodial orientation and the required $U(1)_X$ charge, with $X=(B-L)/2$, we introduce the right-space spurion
\begin{align}
    \mathcal{S}_R\equiv S \sigma_-,\qquad\text{w/ }X(S)=+1
\end{align}
which transforms as
\begin{align}
    \mathcal{S}_R\rightarrow e^{i\alpha_X}g_R\mathcal{S}_R g_R^\dagger
\end{align}
under $SU(2)_L\times SU(2)_R\times U(1)_X$~\footnote{We take $S \sigma_-$ as the right-space LNV spurion, rather than $S\epsilon \sigma_-$, so that the pseudoreality of $SU(2)_L$ is manifest in the dressed left-space object.}. The corresponding dressed left-space spurion is
\begin{align}
    \hat{\mathcal{S}}\equiv U^*\epsilon\mathcal{S}_R U^\dagger~\label{eq:Shat_definition}
\end{align}
with transformation law
\begin{align}
    \hat{\mathcal{S}}\rightarrow e^{i\alpha_X}g_L^*\hat{\mathcal{S}} g_L^\dagger
\end{align}
Thus, $\hat{\mathcal{S}}$ transforms as a symmetric left-left tensor, corresponding to an $SU(2)_L$ triplet, and carries $X(\hat{\mathcal{S}})=+1$, or equivalently $B-L=+2$. The leading LNV HEFT operator can therefore be written in the invariant form
\begin{align}
    \mathcal O_L^{pr}=L_{L p}^T C\hat{\mathcal{S}}L_{L r}
\end{align}
To make sure that the above operator has correct flavor structures, the spurion $\hat{\mathcal{S}}$ should be symmetric in the weak gauge space, which is trivially realized since for the minimal choice $S$ is just a scalar spurion~\footnote{The most general charged right-space spurion $\mathcal{S}_R$ is a $2\times 2$ matrix which can be decomposed in the basis $\{\sigma_+, \sigma_-, \mathbb{1}, \sigma_3\}$. One can easily see that only the $\sigma_-$ direction can select out the $\nu\nu$ component. The presence of Weinberg operator requires $\mathcal{S}_R\propto \sigma_-$, which suggests the minimal choice of the form of $S$ is just a scalar.}. Then the operator is symmetric under $p\leftrightarrow r$ and therefore has $n_f(n_f+1)/2$ independent flavor components.

The spurion $\hat{\mathcal{S}}$ should more precisely be regarded as a $B-L$-violating spurion rather than solely as an LNV spurion. It may therefore also occur in sectors that violate baryon number. Purely baryon-number-violating operators constructed only from SM quarks require at least six fermion fields and occur beyond the chiral orders considered here. We consequently focus on operators that violate lepton number.

Using $\hat{\mathcal{S}}$, together with the usual HEFT building blocks, we construct the $B-L$-violating LNV operators through $\mathcal O(p^4)$. The resulting operators and their flavor multiplicities are listed in TABLE~\ref{tab:LNV_basis}. Operators related by Hermitian conjugation are not listed separately, and undifferentiated powers of $h$ are understood to be absorbed into arbitrary Wilson coefficient functions of $h/v_{\rm EW}$.

The $B-L$-conserving $LQ^3$ class, which violates baryon and lepton number with $\Delta B=\Delta L$, is already contained in the conventional HEFT basis and does not require an insertion of $\hat{\mathcal{S}}$. Conversely, the $\bar LQ^3$ class listed in TABLE~\ref{tab:LNV_basis} violates both baryon and lepton number but carries nonzero $B-L$. Thus, the table contains the newly constructed $B-L$-violating sectors, while the established $LQ^3$ class must be included to obtain the full LNV classification. The operators in TABLE~\ref{tab:LNV_basis}, together with that class and their Hermitian conjugates, constitute the complete LNV HEFT basis through $\mathcal{O}(p^4)$ for the light-field content specified above. 
\setlength{\LTcapwidth}{\textwidth}
\begin{longtable}{|c|c|r|r|}
\caption{$B-L$-violating LNV HEFT operators through $\mathcal{O}(p^4)$, organized by operator class. The final column gives the number of independent flavor structures for $n_f$ fermion generations. Hermitian-conjugate operators are not listed separately, and each operator may be multiplied by an arbitrary function of $h/v_{\rm EW}$.}
\label{tab:LNV_basis}\\
\hline
Class & Label & Operator & Flavor count \\
\hline
\endfirsthead

\hline
Class & Label & Operator & Flavor count \\
\hline
\endhead

\hline
\multicolumn{4}{r}{\textit{continued on next page}}\\
\hline
\endfoot

\hline
\endlastfoot

\hline\hline
\multicolumn{4}{c}{\textbf{${\cal O}(p^2)$}}\\
\hline
\classcount{L^2}{\frac{n_f(n_f+1)}{2}} & ${\cal O}_{L}^{pr}$ & $\displaystyle L_{Lp}^{T} C \widehat{\mathcal S} L_{Lr}$ & $\displaystyle \frac{n_f(n_f+1)}{2}$ \\
\hline

\hline\hline
\multicolumn{4}{c}{\textbf{${\cal O}(p^3)$}}\\
\hline
\classcount{L^2V}{n_f^2} & ${\cal O}_{LV}^{pr}$ & $\displaystyle L_{Lp}^{T} C \widehat{\mathcal S}\gamma^\mu V_\mu U L_{Rr}$ & $\displaystyle n_f^2$ \\
\hline

\hline\hline
\multicolumn{4}{c}{\textbf{${\cal O}(p^4)$}}\\
\hline

\multirow{3}{*}{\classcount{L^2X}{2n_f^2-n_f}}
& ${\cal O}_{LX,B}^{pr}$ & $\displaystyle L_{Lp}^{T} C \widehat{\mathcal S}\sigma_{\mu\nu} L_{Lr} B^{\mu\nu}$ & $\displaystyle \frac{n_f(n_f-1)}{2}$ \\
& ${\cal O}_{LX,W}^{pr}$ & $\displaystyle L_{Lp}^{T} C \widehat{\mathcal S}\sigma_{\mu\nu} W^{\mu\nu} L_{Lr}$ & $\displaystyle n_f^2$ \\
& ${\cal O}_{LX,TW}^{pr}$ & $\displaystyle L_{Lp}^{T} C \widehat{\mathcal S}\sigma_{\mu\nu} L_{Lr}\langle T W^{\mu\nu}\rangle$ & $\displaystyle \frac{n_f(n_f-1)}{2}$ \\
\hline

\multirow{7}{*}{\classcount{L^2V^2}{\frac{9n_f^2+3n_f}{2}}}
& ${\cal O}_{LVV,1}^{pr}$ & $\displaystyle L_{Lp}^{T} C \widehat{\mathcal S} L_{Lr}\langle V_\mu V^\mu\rangle$ & $\displaystyle \frac{n_f(n_f+1)}{2}$ \\
& ${\cal O}_{LVV,2}^{pr}$ & $\displaystyle L_{Lp}^{T} C \widehat{\mathcal S} L_{Lr}\langle T V_\mu\rangle\langle T V^\mu\rangle$ & $\displaystyle \frac{n_f(n_f+1)}{2}$ \\
& ${\cal O}_{LVV,3}^{pr}$ & $\displaystyle L_{Lp}^{T} C \widehat{\mathcal S}V_\mu L_{Lr}\langle T V^\mu\rangle$ & $\displaystyle n_f^2$ \\
& ${\cal O}_{LVV,4}^{pr}$ & $\displaystyle L_{Lp}^{T} C\epsilon\widehat{\mathcal S}^{\dagger}\epsilon L_{Lr}\langle \epsilon\widehat{\mathcal S}V_\mu\rangle\langle \epsilon\widehat{\mathcal S}V^\mu\rangle$ & $\displaystyle \frac{n_f(n_f+1)}{2}$ \\
& ${\cal O}_{LVV,5}^{pr}$ & $\displaystyle L_{Lp}^{T} C \widehat{\mathcal S}\sigma_{\mu\nu}[V^\mu,V^\nu]L_{Lr}$ & $\displaystyle n_f^2$ \\
& ${\cal O}_{LVV,6}^{pr}$ & $\displaystyle L_{Lp}^{T} C \widehat{\mathcal S}\sigma_{\mu\nu}L_{Lr}\langle T[V^\mu,V^\nu]\rangle$ & $\displaystyle \frac{n_f(n_f-1)}{2}$ \\
& ${\cal O}_{LVV,7}^{pr}$ & $\displaystyle L_{Rp}^{T} C U^{T}\epsilon\widehat{\mathcal S}^{\dagger}\epsilon U L_{Rr}\langle \epsilon\widehat{\mathcal S}V_\mu\rangle\langle \epsilon\widehat{\mathcal S}V^\mu\rangle$ & $\displaystyle \frac{n_f(n_f+1)}{2}$ \\
\hline

\classcount{L^2h^2D^2}{\frac{n_f(n_f+1)}{2}} & ${\cal O}_{LhhD^2}^{pr}$ & $\displaystyle L_{Lp}^{T} C\widehat{\mathcal S}L_{Lr}D_\mu h\,D^\mu h$ & $\displaystyle \frac{n_f(n_f+1)}{2}$ \\
\hline

\multirow{4}{*}{\classcount{L^2hDV}{3n_f^2}}
& ${\cal O}_{LhDV,1}^{pr}$ & $\displaystyle L_{Lp}^{T} C\widehat{\mathcal S}V_\mu L_{Lr}\,D^\mu h$ & $\displaystyle n_f^2$ \\
& ${\cal O}_{LhDV,2}^{pr}$ & $\displaystyle L_{Lp}^{T} C\widehat{\mathcal S}L_{Lr}\langle T V_\mu\rangle D^\mu h$ & $\displaystyle \frac{n_f(n_f+1)}{2}$ \\
& ${\cal O}_{LhDV,3}^{pr}$ & $\displaystyle L_{Lp}^{T} C\widehat{\mathcal S}\sigma_{\mu\nu}V^\nu L_{Lr}\,D^\mu h$ & $\displaystyle n_f^2$ \\
& ${\cal O}_{LhDV,4}^{pr}$ & $\displaystyle L_{Lp}^{T} C\widehat{\mathcal S}\sigma_{\mu\nu}L_{Lr}\langle T V^\nu\rangle D^\mu h$ & $\displaystyle \frac{n_f(n_f-1)}{2}$ \\
\hline

\multicolumn{4}{|c|}{\textbf{$\psi^4$ pure Leptonic}}\\
\hline\hline

\classcount{L^4}{\frac{n_f^2(n_f^2-1)}{12}} & ${\cal O}_{L^4}^{prst}$ & $\displaystyle Y\!\left[\begin{smallmatrix}p & r \\ s & t\end{smallmatrix}\right]\big(L_{Lp}^{T}C\widehat{\mathcal S}L_{Lr}\big)\big(L_{Ls}^{T}C\widehat{\mathcal S}L_{Lt}\big)$ & $\displaystyle \frac{n_f^2(n_f^2-1)}{12}$ \\
\hline

\multirow{3}{*}{\classcount{L^3\bar L}{\frac{n_f^2(3n_f^2+n_f)}{2}}}
& ${\cal O}_{L^3\bar L,1}^{prst}$ & $\displaystyle \big(L_{Lp}^{T}C\widehat{\mathcal S}L_{Lr}\big)\big(\bar L_{Ls}U L_{Rt}\big)$ & $\displaystyle \frac{n_f^3(n_f+1)}{2}$ \\
& ${\cal O}_{L^3\bar L,2}^{prst}$ & $\displaystyle \big(L_{Lp}^{T}C\widehat{\mathcal S}L_{Lr}\big)\big(\bar L_{Rs}U^\dagger L_{Lt}\big)$ & $\displaystyle \frac{n_f^3(n_f+1)}{2}$ \\
& ${\cal O}_{L^3\bar L,3}^{prst}$ & $\displaystyle \big(L_{Lp}^{T}C\epsilon L_{Lr}\big)\big(\bar L_{Rs}U^\dagger\epsilon\widehat{\mathcal S}L_{Lt}\big)$ & $\displaystyle \frac{n_f^3(n_f-1)}{2}$ \\
\hline

\multicolumn{4}{|c|}{\textbf{$\psi^4$ semi-Leptonic}}\\
\hline\hline

\multirow{14}{*}{\classcount{L^2\bar Q Q}{n_f^3(8n_f+1)}}
& ${\cal O}_{SQ,LR}^{\widehat S,1\,prst}$ & $\displaystyle \big(L_{Lp}^{T}C\widehat{\mathcal S}L_{Lr}\big)\big(\bar Q_{Ls}UQ_{Rt}\big)$ & $\displaystyle \frac{n_f^3(n_f+1)}{2}$ \\
& ${\cal O}_{SQ,LR}^{\widehat S,T\,prst}$ & $\displaystyle \big(L_{Lp}^{T}C\widehat{\mathcal S}L_{Lr}\big)\big(\bar Q_{Ls}TUQ_{Rt}\big)$ & $\displaystyle \frac{n_f^3(n_f+1)}{2}$ \\
& ${\cal O}_{SQ,RL}^{\widehat S,1\,prst}$ & $\displaystyle \big(L_{Lp}^{T}C\widehat{\mathcal S}L_{Lr}\big)\big(\bar Q_{Rs}U^\dagger Q_{Lt}\big)$ & $\displaystyle \frac{n_f^3(n_f+1)}{2}$ \\
& ${\cal O}_{SQ,RL}^{\widehat S,T\,prst}$ & $\displaystyle \big(L_{Lp}^{T}C\widehat{\mathcal S}L_{Lr}\big)\big(\bar Q_{Rs}U^\dagger TQ_{Lt}\big)$ & $\displaystyle \frac{n_f^3(n_f+1)}{2}$ \\
& ${\cal O}_{TQ,RL}^{\widehat S,1\,prst}$ & $\displaystyle \big(L_{Lp}^{T}C\sigma_{\mu\nu}\widehat{\mathcal S}L_{Lr}\big)\big(\bar Q_{Rs}\sigma^{\mu\nu}U^\dagger Q_{Lt}\big)$ & $\displaystyle \frac{n_f^3(n_f-1)}{2}$ \\
& ${\cal O}_{TQ,RL}^{\widehat S,T\,prst}$ & $\displaystyle \big(L_{Lp}^{T}C\sigma_{\mu\nu}\widehat{\mathcal S}L_{Lr}\big)\big(\bar Q_{Rs}\sigma^{\mu\nu}U^\dagger TQ_{Lt}\big)$ & $\displaystyle \frac{n_f^3(n_f-1)}{2}$ \\
\cline{2-4}
& ${\cal O}_{SQ,LR}^{\epsilon\,prst}$ & $\displaystyle \big(L_{Lp}^{T}C\epsilon L_{Lr}\big)\big(\bar Q_{Ls}\epsilon\widehat{\mathcal S}UQ_{Rt}\big)$ & $\displaystyle \frac{n_f^3(n_f-1)}{2}$ \\
& ${\cal O}_{SQ,LR}^{\epsilon T\,prst}$ & $\displaystyle \big(L_{Lp}^{T}C\epsilon T L_{Lr}\big)\big(\bar Q_{Ls}\epsilon\widehat{\mathcal S}UQ_{Rt}\big)$ & $\displaystyle \frac{n_f^3(n_f+1)}{2}$ \\
& ${\cal O}_{SQ,RL}^{\epsilon\,prst}$ & $\displaystyle \big(L_{Lp}^{T}C\epsilon L_{Lr}\big)\big(\bar Q_{Rs}U^\dagger\epsilon\widehat{\mathcal S}Q_{Lt}\big)$ & $\displaystyle \frac{n_f^3(n_f-1)}{2}$ \\
& ${\cal O}_{SQ,RL}^{\epsilon T\,prst}$ & $\displaystyle \big(L_{Lp}^{T}C\epsilon T L_{Lr}\big)\big(\bar Q_{Rs}U^\dagger\epsilon\widehat{\mathcal S}Q_{Lt}\big)$ & $\displaystyle \frac{n_f^3(n_f+1)}{2}$ \\
& ${\cal O}_{TQ,RL}^{\epsilon\,prst}$ & $\displaystyle \big(L_{Lp}^{T}C\sigma_{\mu\nu}\epsilon L_{Lr}\big)\big(\bar Q_{Rs}\sigma^{\mu\nu}U^\dagger\epsilon\widehat{\mathcal S}Q_{Lt}\big)$ & $\displaystyle \frac{n_f^3(n_f-1)}{2}$ \\
& ${\cal O}_{TQ,RL}^{\epsilon T\,prst}$ & $\displaystyle \big(L_{Lp}^{T}C\sigma_{\mu\nu}\epsilon T L_{Lr}\big)\big(\bar Q_{Rs}\sigma^{\mu\nu}U^\dagger\epsilon\widehat{\mathcal S}Q_{Lt}\big)$ & $\displaystyle \frac{n_f^3(n_f+1)}{2}$ \\
& ${\cal O}_{VQ,L}^{prst}$ & $\displaystyle \big(L_{Lp}^{T}C\gamma_\mu\epsilon U L_{Rr}\big)\big(\bar Q_{Ls}\gamma^\mu\epsilon\widehat{\mathcal S}Q_{Lt}\big)$ & $\displaystyle n_f^4$ \\
& ${\cal O}_{VQ,R}^{prst}$ & $\displaystyle \big(L_{Lp}^{T}C\gamma_\mu\epsilon U L_{Rr}\big)\big(\bar Q_{Rs}U^\dagger\gamma^\mu\epsilon\widehat{\mathcal S}UQ_{Rt}\big)$ & $\displaystyle n_f^4$ \\
\hline

\multirow{9}{*}{\classcount{\bar L Q^3}{\frac{n_f^2(25n_f^2-9n_f-4)}{6}}}
& ${\cal O}_{A1}^{L;\bar{L}R\,prst}$ & $\displaystyle \epsilon_{\alpha\beta\gamma}\big(Q_{Lp}^{\alpha T}C\epsilon Q_{Lr}^{\beta}\big)\big(\bar L_{Ls}\widehat{\mathcal S}^{\dagger}\epsilon UQ_{Rt}^{\gamma}\big)$ & $\displaystyle \frac{n_f^3(n_f+1)}{2}$ \\
& ${\cal O}_{A2}^{L;\bar{L}R\,prst}$ & $\displaystyle \epsilon_{\alpha\beta\gamma}\big(Q_{Lp}^{\alpha T}C\epsilon TQ_{Lr}^{\beta}\big)\big(\bar L_{Ls}\widehat{\mathcal S}^{\dagger}\epsilon UQ_{Rt}^{\gamma}\big)$ & $\displaystyle \frac{n_f^3(n_f-1)}{2}$ \\
& ${\cal O}_{B1}^{L;\bar{L}R\,prst}$ & $\displaystyle \epsilon_{\alpha\beta\gamma}\big(Q_{Lp}^{\alpha T}C\epsilon\widehat{\mathcal S}^{\dagger}\epsilon Q_{Lr}^{\beta}\big)\big(\bar L_{Ls}UQ_{Rt}^{\gamma}\big)$ & $\displaystyle \frac{n_f^3(n_f-1)}{2}$ \\
& ${\cal O}_{B2}^{L;\bar{L}R\,prst}$ & $\displaystyle \epsilon_{\alpha\beta\gamma}\big(Q_{Lp}^{\alpha T}C\epsilon\widehat{\mathcal S}^{\dagger}\epsilon Q_{Lr}^{\beta}\big)\big(\bar L_{Ls}TUQ_{Rt}^{\gamma}\big)$ & $\displaystyle \frac{n_f^3(n_f-1)}{2}$ \\
& ${\cal O}_{B}^{L;\bar{R}L\,prst}$ & $\displaystyle Y\!\left[\begin{smallmatrix}p & t \\ r & {}\end{smallmatrix}\right]\epsilon_{\alpha\beta\gamma}\big(Q_{Lp}^{\alpha T}C\epsilon\widehat{\mathcal S}^{\dagger}\epsilon Q_{Lr}^{\beta}\big)\big(\bar L_{Rs}U^\dagger Q_{Lt}^{\gamma}\big)$ & $\displaystyle \frac{n_f^2(n_f^2-1)}{3}$ \\
& ${\cal O}_{B}^{R;\bar{R}L\,prst}$ & $\displaystyle \epsilon_{\alpha\beta\gamma}\big(Q_{Rp}^{\alpha T}CU^T\epsilon\widehat{\mathcal S}^{\dagger}\epsilon UQ_{Rr}^{\beta}\big)\big(\bar L_{Rs}U^\dagger Q_{Lt}^{\gamma}\big)$ & $\displaystyle \frac{n_f^3(n_f-1)}{2}$ \\
& ${\cal O}_{A1}^{R;\bar{L}R\,prst}$ & $\displaystyle Y[p\,r]\epsilon_{\alpha\beta\gamma}\big(Q_{Rp}^{\alpha T}CU^T\epsilon UQ_{Rr}^{\beta}\big)\big(\bar L_{Ls}\widehat{\mathcal S}^{\dagger}\epsilon UQ_{Rt}^{\gamma}\big)$ & $\displaystyle \frac{n_f^3(n_f+1)}{2}$ \\
& ${\cal O}_{A2}^{R;\bar{L}R\,prst}$ & $\displaystyle Y[p\,r]\epsilon_{\alpha\beta\gamma}\big(Q_{Rp}^{\alpha T}CU^T\epsilon T UQ_{Rr}^{\beta}\big)\big(\bar L_{Ls}\widehat{\mathcal S}^{\dagger}\epsilon UQ_{Rt}^{\gamma}\big)$ & $\displaystyle \frac{n_f^3(n_f-1)}{2}$ \\
& ${\cal O}_{B-}^{R;\bar{L}R\,prst}$ & \makecell[r]{$\displaystyle Y\!\left[\begin{smallmatrix}p & t \\ r & {}\end{smallmatrix}\right]\epsilon_{\alpha\beta\gamma}\big(Q_{Rp}^{\alpha T}CU^T\epsilon\widehat{\mathcal S}^{\dagger}\epsilon UQ_{Rr}^{\beta}\big)$ \\ $\Big[\big(\bar L_{Ls}UQ_{Rt}^{\gamma}\big)-\big(\bar L_{Ls}TUQ_{Rt}^{\gamma}\big)\Big]$} & $\displaystyle \frac{n_f^2(n_f^2-1)}{3}$ \\
\hline
\end{longtable}

The construction of the $\bar{L}Q^3$ class requires some additional care. We divide its operators into two subclasses. In the type-A operators, $\hat{\mathcal{S}}^\dagger$ appears in the $\bar{L}Q$ bilinear, whereas in the type-B operators it appears in the $QQ$ bilinear. One might initially introduce two type-B structures, $\mathcal{O}^{R; \bar{L}R}_{B1}=\epsilon_{\alpha\beta\gamma}(Q_{R p}^{\alpha T}C U^T\epsilon\hat{\mathcal{S}}^\dagger\epsilon U Q_{R r}^\beta)(\bar{L}_{L s}U Q_{R t}^\gamma)$ and $\mathcal{O}^{R; \bar{L}R}_{B2}=\epsilon_{\alpha\beta\gamma}(Q_{R p}^{\alpha T}C U^T\epsilon\hat{\mathcal{S}}^\dagger\epsilon U Q_{R r}^\beta)(\bar{L}_{L s}T U Q_{R t}^\gamma)$. However, the combination $\mathcal{O}_{B1}^{R;\bar{L}R}+\mathcal{O}_{B2}^{R;\bar{L}R}$ lies in the linear span of the two type-A operators $\mathcal{O}_{A1}^{R;\bar{L}R}$ and $\mathcal{O}_{A2}^{R;\bar{L}R}$. The corresponding relation follows from a nontrivial spurion identity and is derived explicitly in App.~\ref{app:TypeBred}. Consequently, the two type-B monomials do not define two independent operator directions; only one linear combination is independent. This is why the mixed $(\widehat S,T)$ representation of this class contains a nonmonomial basis element. A purely $(\hat{\mathcal{S}},\hat{\mathcal{S}}^\dagger)$ representation restores a monomial choice, as shown below.

\subsection{LNV and custodial-symmetry breaking}
The simultaneous use of the LNV spurion $\hat{\mathcal{S}}$ and the custodial spurion $T$ provides a transparent classification of the operators, but the two spurions are not algebraically independent. For the minimal orientation defined in Eq.~\eqref{eq:Shat_definition}, one finds
\begin{align}
    \hat{\mathcal{S}}^\dagger\hat{\mathcal{S}}=\frac{|S|^2}{2}(\mathbb{1}+T)=\frac{1}{2}(\mathbb{1}+T)~\label{eq:SdagS_T_relation}. 
\end{align}
Thus, for a fixed nonzero spurion background, insertions of $T$ can be rewritten in terms of neutral combinations of $\hat{\mathcal{S}}$ and $\hat{\mathcal{S}}^\dagger$. The operator basis may therefore be expressed entirely in terms of the LNV spurion and its Hermitian conjugate. For the $\bar{L}Q^3$ class, the resulting pure-$(\hat{\mathcal{S}},\hat{\mathcal{S}}^\dagger)$ basis is shown in TABLE~\ref{tab:pureS_LbarQ3}.
\begin{longtable}{lrr}
\caption{Pure-$(\hat{\mathcal{S}},\hat{\mathcal{S}}^\dagger)$ representation of the $\bar{L}Q^3$ operator class.}
\label{tab:pureS_LbarQ3}\\
\hline
Label & Operator & Flavor count \\
\hline
\endfirsthead

\hline
Label & Operator & Flavor count \\
\hline
\endhead

\hline
\multicolumn{3}{r}{\textit{continued on next page}}\\
\hline
\endfoot

\hline
\endlastfoot

${\cal O}_{A}^{L;\bar{L}R\,prst}$
& $\displaystyle \epsilon_{\alpha\beta\gamma}\big(Q_{Lp}^{\alpha T}C\epsilon\widehat{\mathcal S}^{\dagger}\widehat{\mathcal S}Q_{Lr}^{\beta}\big)\big(\bar L_{Ls}\widehat{\mathcal S}^{\dagger}\epsilon UQ_{Rt}^{\gamma}\big)$
& $\displaystyle n_f^4$ \\

${\cal O}_{B1}^{L;\bar{L}R\,prst}$
& $\displaystyle \epsilon_{\alpha\beta\gamma}\big(Q_{Lp}^{\alpha T}C\epsilon\widehat{\mathcal S}^{\dagger}\epsilon Q_{Lr}^{\beta}\big)\big(\bar L_{Ls}\widehat{\mathcal S}^{\dagger}\widehat{\mathcal S}UQ_{Rt}^{\gamma}\big)$
& $\displaystyle \frac{n_f^3(n_f-1)}{2}$ \\

${\cal O}_{B2}^{L;\bar{L}R\,prst}$
& $\displaystyle \epsilon_{\alpha\beta\gamma}\big(Q_{Lp}^{\alpha T}C\epsilon\widehat{\mathcal S}^{\dagger}\epsilon Q_{Lr}^{\beta}\big)\big(\bar L_{Ls}\epsilon\widehat{\mathcal S}\widehat{\mathcal S}^{\dagger}\epsilon UQ_{Rt}^{\gamma}\big)$
& $\displaystyle \frac{n_f^3(n_f-1)}{2}$ \\

${\cal O}_{B}^{L;\bar{R}L\,prst}$
& $\displaystyle Y\!\left[\begin{smallmatrix} p & t\\ r & \phantom{t} \end{smallmatrix}\right]\epsilon_{\alpha\beta\gamma}\big(Q_{Lp}^{\alpha T}C\epsilon\widehat{\mathcal S}^{\dagger}\epsilon Q_{Lr}^{\beta}\big)\big(\bar L_{Rs}U^\dagger Q_{Lt}^{\gamma}\big)$
& $\displaystyle \frac{n_f^2(n_f^2-1)}{3}$ \\

${\cal O}_{B}^{R;\bar{R}L\,prst}$
& $\displaystyle \epsilon_{\alpha\beta\gamma}\big(Q_{Rp}^{\alpha T}CU^T\epsilon\widehat{\mathcal S}^{\dagger}\epsilon UQ_{Rr}^{\beta}\big)\big(\bar L_{Rs}U^\dagger Q_{Lt}^{\gamma}\big)$
& $\displaystyle \frac{n_f^3(n_f-1)}{2}$ \\

${\cal O}_{A}^{R;\bar{L}R\,prst}$
& $\displaystyle \epsilon_{\alpha\beta\gamma}\big(Q_{Rp}^{\alpha T}CU^T\epsilon\widehat{\mathcal S}^{\dagger}\widehat{\mathcal S}UQ_{Rr}^{\beta}\big)\big(\bar L_{Ls}\widehat{\mathcal S}^{\dagger}\epsilon UQ_{Rt}^{\gamma}\big)$
& $\displaystyle n_f^4$ \\

${\cal O}_{B}^{R;\bar{L}R\,prst}$
& $\displaystyle Y\!\left[\begin{smallmatrix} p & t\\ r & \phantom{t} \end{smallmatrix}\right]\epsilon_{\alpha\beta\gamma}\big(Q_{Rp}^{\alpha T}CU^T\epsilon\widehat{\mathcal S}^{\dagger}\epsilon UQ_{Rr}^{\beta}\big)\big(\bar L_{Ls}\epsilon\widehat{\mathcal S}\widehat{\mathcal S}^{\dagger}\epsilon UQ_{Rt}^{\gamma}\big)$
& $\displaystyle \frac{n_f^2(n_f^2-1)}{3}$ \\
\hline
\end{longtable}

Although a pure-$\hat{\mathcal{S}}$ basis is algebraically possible, it obscures the custodial properties of individual operators and generally increases the number of explicit LNV-spurion insertions. We therefore retain both $\hat{\mathcal{S}}$ and $T$ in the operator notation used in the remainder of this work.

The electrically neutral LNV insertion $\hat{\mathcal{S}}$ selects the definite right-space direction $\sigma_-$. It therefore carries a nontrivial custodial orientation. Eq.~\eqref{eq:SdagS_T_relation} further shows that neutral quadratic combinations of LNV spurions contain both a custodial singlet and the custodial-breaking direction $T$. Consequently, within the minimal HEFT containing only the SM light degrees of freedom and a single neutral LNV orientation, the LNV sector is intrinsically accompanied by custodial-symmetry-breaking spurion structures. A custodial symmetric construction would require the LNV sources to form a complete custodial multiplet. For example, introducing
\begin{align}
    \hat{\mathcal{S}}_a\sim U^*\epsilon \sigma_a U^\dagger,
\end{align}
would give
\begin{align}
    \sum_a\hat{\mathcal{S}}_a^\dagger\hat{\mathcal{S}}_a\propto U\left(\sum_a\sigma_a\sigma_a\right)U^\dagger=3\mathbb{1}.
\end{align}
However, the custodial partners of the electrically neutral Majorana insertion carry nonzero electric charge. Such a complete multiplet is absent in the minimal HEFT with only SM light fields and would require additional charged sources or an enlarged low-energy field content.

In the type-II seesaw model, the LNV interaction $\mu\mathcal{H}^{T}i\sigma_{2}\Xi^{\dagger}\mathcal{H}+{\rm h.c.}$ selects precisely the custodial orientation encoded by $\hat{\mathcal{S}}$. Thus, lepton-number violation and the selection of a custodial direction arise from the same renormalizable interaction. By contrast, in the type-I seesaw model, the Majorana mass $M_N$ of the electroweak-singlet fermion violates lepton number but is itself a custodial singlet. The custodial orientation is instead selected by the neutrino Yukawa interaction $y_N\bar{L}_L\widetilde{\mathcal H}N$, which singles out a definite direction in $SU(2)_R$ space. After the heavy fermion $N$ is integrated out, two insertions of the neutrino Yukawa interaction, together with the inverse Majorana mass, generate the Weinberg operator at tree level, with its custodial orientation encoded by $\hat{\mathcal S}$. Therefore, in the type-I seesaw model, lepton-number violation and custodial-symmetry breaking originate from distinct UV spurions, $M_N$ and $y_N$, respectively, although they combine already at tree level in the low-energy LNV operator. This comparison also illustrates why the dressed LNV spurion is an EFT construction: the same low-energy custodial orientation can emerge from different microscopic sources of lepton-number violation.

\subsection{Comparison with previous results and Hilbert-series validation}
We perform two independent checks of the LNV operator basis constructed above. First, after electroweak symmetry breaking, we decompose the four-fermion HEFT operators into their component-field expressions. The resulting operators can be placed in one-to-one correspondence with the complete LNV four-fermion LEFT basis of Ref.~\cite{Jenkins:2017jig}. This correspondence provides an independent check of the Lorentz structures, flavor symmetries, and field content of our four-fermion operator basis.

Second, we compare our operator counting with the HEFT enumeration of Ref.~\cite{Graf:2022rco}. The two analyses employ different methods. Reference~\cite{Graf:2022rco} performs the counting using Hilbert-series techniques, both through a direct construction in terms of the massive low-energy degrees of freedom and through a custodial-spurion formulation. As emphasized in Ref.~\cite{Graf:2022rco}, the custodial-$T$-spurion formulation is equivalent to the direct counting in the $B-L$-conserving sectors, because the unbroken custodial $U(1)_V$ coincides with electromagnetism only when $B-L=0$. By contrast, in the present work we explicitly construct the LNV operators using the complex spurion $\hat{\mathcal{S}}$ and reduce them to a minimal basis by solving the algebraic relations among their component expressions.

Our component-level reduction establishes the linear independence of the operators retained in the basis. Moreover, the number of operators in each field class, including their full flavor multiplicities, agrees with the corresponding Hilbert-series counting of Ref.~\cite{Graf:2022rco}. Since the Hilbert series gives the dimension of the operator space after redundancies associated with integration by parts, equations of motion, and algebraic identities have been removed, the agreement between this dimension and the number of explicitly constructed independent operators establishes both the completeness and the nonredundancy of our LNV HEFT basis through $\mathcal{O}(p^4)$.

To retain explicitly the information carried by the LNV spurion, we also construct a spurion-refined Hilbert series in terms of the complex spurions $\hat{\mathcal{S}}$ and $\hat{\mathcal{S}}^{\dagger}$. The $B-L$-violating fermionic sectors, which provide the direct counting check of the new LNV basis, are summarized in TABLE~\ref{tab:HS_BLneq0}. For completeness, the bosonic and $B-L$-conserving results are also shown in Tables~\ref{tab:HS_bosonic} and~\ref{tab:HS_BL0} below. In each case we also display the corresponding $T$-spurion counting of Ref.~\cite{Graf:2022rco}. The two refinements organize the operator space differently: the $(\hat{\mathcal{S}},\hat{\mathcal{S}}^{\dagger})$ grading keeps track of the $B-L$ charge and the number of LNV-spurion insertions, whereas the $T$ grading resolves the corresponding custodial-symmetry-breaking structures.

\begin{table}[h!]
\centering\small
\begin{tabular}{ccccc}
\toprule
\multicolumn{2}{c}{Class} & dim & $T$ Spurion & $\hat{\mathcal{S}}$ Spurion \\ \midrule
\multirow{3}{*}{$\psi^2 D^2$} & $L^2 V^2 + \text{h.c.}$ & 5 & \makecell{$\frac{2n_f^2-3n_f}{2} + \frac{5n_f^2+3n_f}{2}T$ \\ $+\frac{7n_f^2-3n_f}{4}T^2+\frac{3n_f^2+3n_f}{4}T^3$} & \makecell{$\frac{7n_f^2+n_f}{2}\hat{\mathcal{S}}$ \\ $+(n_f^2+n_f)\hat{\mathcal{S}}^2\hat{\mathcal{S}}^\dagger$} \\
\cmidrule(){2-5}
                              & $L^2 h {\cal D} V + \text{h.c.}$   & 6 & $\frac12(3+4T+3T^2)n_f^2$ & $3n_f^2\hat{\mathcal{S}}$ \\
\cmidrule(){2-5}
                              & $L^2 h^2 {\cal D}^2 + \text{h.c.}$ & 7 & $\frac{n_f^2-3n_f}{4}+\frac{3n_f^2+3n_f}{4}T$ & $\frac12(n_f^2+n_f)\hat{\mathcal{S}}$ \\
\midrule
$\psi^2 D$ & $L^2 V + \text{h.c.}$         & 4 & $\frac12(1+2T+T^2)n_f^2$ & $n_f^2\hat{\mathcal{S}}$ \\
\midrule
$\psi^2 X$ & $L^2 X + \text{h.c.}$ & 5 & \makecell{$n_f^2+\frac14(7n_f^2-3n_f)T$ \\ $+\frac14(3n_f^2-3n_f)T^2$} & \makecell{$(2n_f^2-n_f)\hat{\mathcal{S}}$} \\
\midrule
                          \multirow{4}{*}{$\psi^4$} & $L^4 + \text{h.c.}$ & 6 & \makecell{$\frac{1}{12}n_f^2(2n_f^2+3n_f+13)$ \\ $+\frac13n_f^2(n_f^2-1)T$ \\ $+\frac14n_f^2(n_f^2+n_f)T^2$} & \makecell{$\frac{1}{12}n_f^2(n_f^2-1)\hat{\mathcal{S}}^2$} \\
\cmidrule(){2-5}
                          & $L^3\bar{L} + \text{h.c.}$ & 6 & \makecell{$\frac{1}{12}n_f^2(11n_f^2+3n_f+4)$ \\ $+\frac12n_f^2(3n_f^2+n_f)T$ \\ $+\frac{1}{12}n_f^2(7n_f^2+3n_f-4)T^2$} & \makecell{$\frac{1}{2}n_f^3(3n_f+1)\hat{\mathcal{S}}$} \\
\cmidrule(){2-5}
& $L^2Q\bar{Q} + \text{h.c.}$ & 6 & \makecell{$5n_f^4+\frac14n_f^2(33n_f^2+3n_f)T$ \\ $+\frac14n_f^2(13n_f^2+3n_f)T^2$} & \makecell{$n_f^3(8n_f+1)\hat{\mathcal{S}}$} \\
\cmidrule(){2-5}
                          & $L\bar{Q}^3 + \text{h.c.}$ & 6 & \makecell{$\frac12n_f^2(5n_f^2+1)$ \\ $+\frac14n_f^2(15n_f^2-3n_f)T$ \\ $+\frac14n_f^2(5n_f^2-3n_f-2)T^2$} & \makecell{$\frac16n_f^2(25n_f^2-9n_f-4)\hat{\mathcal{S}}$} \\
\bottomrule
\end{tabular}
\caption{Spurion-refined Hilbert-series results for fermionic $B-L$-violating HEFT operator classes through NLO.}
\label{tab:HS_BLneq0}
\end{table}

The corresponding bosonic and $B-L$-conserving operator classes are collected in Tables~\ref{tab:HS_bosonic} and~\ref{tab:HS_BL0}. These neutral sectors provide a useful comparison of the two refinements without introducing the extra $B-L$ grading of the LNV classes.

\begin{table}[htbp]
\centering
\begin{tabular}{ccccc}
\toprule
Class & Detailed Class  & dim & $T$ Spurion & $\hat{\mathcal{S}}$ Spurion \\ \midrule
\multirow{5}{*}{$D^4$} & $V^4$           & 4 & $2+2T^2+T^4$ & $2+2\hat{\mathcal{S}}\hat{\mathcal{S}}^\dagger+\hat{\mathcal{S}}^2\hat{\mathcal{S}}^{\dagger 2}$ \\
                       & $h {\cal D} V^3$     & 5 & $2T+T^2+T^3$ & $3\hat{\mathcal{S}}\hat{\mathcal{S}}^\dagger+\hat{\mathcal{S}}^2\hat{\mathcal{S}}^{\dagger 2}$ \\
                       & $h^2 {\cal D}^2 V^2$ & 6 & $2+2T^2$ & $2+2\hat{\mathcal{S}}\hat{\mathcal{S}}^\dagger$ \\
                       & $h^3 {\cal D}^3 V$   & 7 & $T$ & $\hat{\mathcal{S}}\hat{\mathcal{S}}^\dagger$ \\
                       & $h^4 {\cal D}^4$     & 8 & $1$ & $1$ \\
\midrule
$D^2 X$ & $V^2 X$      & 4 & $2+4T+2T^2$ & $2+6\hat{\mathcal{S}}\hat{\mathcal{S}}^\dagger$ \\
\midrule
$X^2$ &                & 4 & $6+2T+2T^2$ & $6+4\hat{\mathcal{S}}\hat{\mathcal{S}}^\dagger$ \\
\midrule
$X^3$ &                & 6 & $4+2T$ & $4+2\hat{\mathcal{S}}\hat{\mathcal{S}}^\dagger$ \\
\bottomrule
\end{tabular}
\caption{Spurion-refined Hilbert-series results for the bosonic HEFT operator classes through NLO.}
\label{tab:HS_bosonic}
\end{table}

\begin{table}[htbp]
\centering\small
\begin{tabular}{ccccc}
\toprule
\multicolumn{2}{c}{Class} & dim & $T$ Spurion & $\hat{\mathcal{S}}$ Spurion \\ \midrule
\multirow{6}{*}{$\psi^2 D^2$} & $L\bar{L} V^2$       & 5 & $(2+4T+3T^2+T^3) n_f^2$  & $(2+7\hat{\mathcal{S}}\hat{\mathcal{S}}^\dagger+\hat{\mathcal{S}}^2\hat{\mathcal{S}}^{\dagger 2}) n_f^2$ \\
                              & $L\bar{L} h {\cal D} V$   & 6 & $(2+4T+2T^2) n_f^2$      & $(2+6\hat{\mathcal{S}}\hat{\mathcal{S}}^\dagger) n_f^2$ \\
                              & $L\bar{L} h^2 {\cal D}^2$ & 7 & $(1+T) n_f^2$            & $(1+1\hat{\mathcal{S}}\hat{\mathcal{S}}^\dagger) n_f^2$ \\
                              & $Q\bar{Q} V^2$       & 5 & $(4+8T+6T^2+2T^3) n_f^2$ & $(4+14\hat{\mathcal{S}}\hat{\mathcal{S}}^\dagger+2\hat{\mathcal{S}}^2\hat{\mathcal{S}}^{\dagger 2}) n_f^2$ \\
                              & $Q\bar{Q} h {\cal D} V$   & 6 & $(4+8T+4T^2) n_f^2$      & $(4+12\hat{\mathcal{S}}\hat{\mathcal{S}}^\dagger) n_f^2$ \\
                              & $Q\bar{Q} h^2 {\cal D}^2$ & 7 & $(2+2T) n_f^2$ & $(2+2\hat{\mathcal{S}}\hat{\mathcal{S}}^\dagger) n_f^2$ \\
\midrule
\multirow{2}{*}{$\psi^2 D$} & $L\bar{L} V$ & 4 & $(1+3T+ T^2) n_f^2$ & $(1+4\hat{\mathcal{S}}\hat{\mathcal{S}}^\dagger) n_f^2$ \\
                            & $Q\bar{Q} V$ & 4 & $(2+4T+2T^2) n_f^2$ & $(2+6\hat{\mathcal{S}}\hat{\mathcal{S}}^\dagger) n_f^2$ \\
\midrule
\multirow{2}{*}{$\psi^2 X$} & $L\bar{L} X$ & 5 & $(2+3T+ T^2) n_f^2$ & $(2+4\hat{\mathcal{S}}\hat{\mathcal{S}}^\dagger) n_f^2$ \\
                            & $Q\bar{Q} X$ & 5 & $(6+8T+2T^2) n_f^2$ & $(6+10\hat{\mathcal{S}}\hat{\mathcal{S}}^\dagger) n_f^2$ \\
\midrule
         & $(L\bar{L})^2$ & 6 & \makecell{$\frac14 n_f^2 (7n_f^2+2n_f+7)$ \\ $+\frac14 n_f^2 (9n_f^2+2n_f-3) T$ \\ $+\frac14 n_f^2 (3n_f^2+2n_f+3) T^2$} & \makecell{$\frac14 n_f^2 (7n_f^2+2n_f+7)$ \\ $n_f^3(3n_f+1)\hat{\mathcal{S}}\hat{\mathcal{S}}^\dagger$} \\
\cmidrule(){2-5}
$\psi^4$ & $(L\bar{L})(Q\bar{Q})$ & 6 & $(12+17T+5T^2) n_f^4$ & $(12+22\hat{\mathcal{S}}\hat{\mathcal{S}}^\dagger) n_f^4$ \\
\cmidrule(){2-5}
         & $(Q\bar{Q})^2$ & 6 & \makecell{$n_f^2 (10n_f^2+6)$ \\ $+n_f^2 (15n_f^2-3) T$ \\ $+n_f^2 ( 5n_f^2+3) T^2$} & \makecell{$n_f^2(10n_f^2+6)$ \\ $20n_f^4\hat{\mathcal{S}}\hat{\mathcal{S}}^\dagger$} \\
\cmidrule(){2-5}
         & $LQ^3 + \text{h.c.}$ & 6 & \makecell{$\frac12n_f^2(5n_f^2+1)$ \\ $+\frac14n_f^2(15n_f^2-3n_f)T$ \\ $+\frac14n_f^2(5n_f^2-3n_f-2)T^2$} & \makecell{$\frac13n_f^2(5n_f^2+1)$ \\ $+\frac13n_f^2(10n_f^2-3n_f-1)\hat{\mathcal{S}}\hat{\mathcal{S}}^\dagger$ \\ $\frac12n_f^3(5n_f-1)\hat{\mathcal{S}}^2(\hat{\mathcal{S}}^{\dagger})^2$} \\
\bottomrule
\end{tabular}
\caption{Spurion-refined Hilbert-series results for fermionic $B-L$-conserving HEFT operator classes through NLO.}
\label{tab:HS_BL0}
\end{table}

In the $B-L$-conserving sectors, the $\hat{\mathcal{S}}$-spurion result can be mapped onto the conventional $T$-spurion counting by treating the neutral combinations $\hat{\mathcal{S}}^{n}(\hat{\mathcal{S}}^{\dagger})^{n}$, where $n$ denotes their common power, as custodial tensors and decomposing them into irreducible representations of $SU(2)_V$. The basic neutral pair transforms as the product of two custodial triplets,
\begin{align}
    \mathbf{3}_{\hat{\mathcal{S}}}\otimes\mathbf{3}_{\hat{\mathcal{S}}^\dagger}=\mathbf{1}\oplus\mathbf{3}\oplus\mathbf{5}
\end{align}
The singlet component is a pure-spurion invariant and is removed by the usual spurion quotient, whereas the nonsinglet components are rewritten in the reduced $T$-spurion basis. In particular, the isospin-one component is mapped onto a single $T$ insertion, while the isospin-two component is mapped onto the symmetric-traceless two-$T$ structure,
\begin{align}
    \mathbf{3}\leftrightarrow T,\qquad\mathbf{5}\leftrightarrow(TT)_{\rm ST}\sim T^2
\end{align}

More generally, for each neutral sector $\hat{\mathcal{S}}^{n}(\hat{\mathcal S}^{\dagger})^{n}$, we decompose the appropriate symmetrized spurion tensor products into irreducible representations of $SU(2)_V$, remove pure-spurion singlet towers and other redundant components using the spurion identities, and rewrite the surviving isospin-$I$ components in the reduced $T$-spurion basis. This reduction is operator-class dependent because the tensor structure of the light fields determines which custodial-isospin components can form independent invariants.

For example, in the $V^4$ class, the $\hat{\mathcal{S}}$-refined Hilbert series
\begin{align}
    H_S(V^4)=2+2\hat{\mathcal{S}}\hat{\mathcal{S}}^\dagger+\hat{\mathcal{S}}^2\hat{\mathcal{S}}^{\dagger 2}
\end{align}
reduces to
\begin{align}
    H_T(V^4)=2+2T^2+T^4,
\end{align}
because the isospin-one component of $\hat{\mathcal{S}}\hat{\mathcal{S}}^{\dagger}$ does not generate an independent invariant in this class. By contrast, in the $h\mathcal{D}V^3$ class, the same neutral spurion pair contributes through both its isospin-one and isospin-two components. Accordingly,
\begin{align}
    H_S(h\mathcal{D}V^3)=3\hat{\mathcal{S}}\hat{\mathcal{S}}^\dagger+\hat{\mathcal{S}}^2\hat{\mathcal{S}}^{\dagger 2}
\end{align}
is reduced according to
\begin{align}
    3\hat{\mathcal{S}}\hat{\mathcal{S}}^\dagger\rightarrow 2T+T^2,\qquad\hat{\mathcal{S}}^2\hat{\mathcal{S}}^{\dagger 2}\rightarrow T^3,
\end{align}
yielding
\begin{align}
    H_T(h\mathcal{D}V^3)=2T+T^2+T^3.
\end{align}
These examples illustrate that the $\hat{\mathcal{S}}$ grading counts neutral combinations of $\hat{\mathcal{S}}$ and $\hat{\mathcal{S}}^{\dagger}$, whereas the $T$ grading resolves the corresponding reduced custodial-isospin structures. After the appropriate representation-theoretic reduction, setting the remaining spurion variables to unity yields the same unrefined operator count as the conventional $T$-spurion method in the corresponding $B-L$-conserving sector.

In the $B-L$-violating sectors, the Hilbert series constructed with $\hat{\mathcal{S}}$ and $\hat{\mathcal{S}}^{\dagger}$ retains the $B-L$ grading and yields the spurion structures and total numbers of independent operators summarized in TABLE~\ref{tab:HS_BLneq0}. Upon setting the spurion variables to unity, the total numbers agree with the direct massive-particle counting of Ref.~\cite{Graf:2022rco}. We also display the corresponding $T$-spurion results for comparison. Since both $T$ and $\hat{\mathcal S}$ transform as custodial-isospin triplets, one may expect a formal representation-theoretic relation between the two countings after imposing $\hat{\mathcal{S}}^{\dagger}=\hat{\mathcal{S}}$. Such an identification, however, discards the $B-L$ grading: $T$ is neutral under $B-L$, whereas $\hat{\mathcal{S}}$ and $\hat{\mathcal{S}}^{\dagger}$ carry opposite $B-L$ charges. Consequently, the physical interpretation of the $T$-spurion counting in the $B-L$-violating sectors, as well as its precise correspondence with the $(\hat{\mathcal{S}},\hat{\mathcal{S}}^{\dagger})$ refinement, remains to be clarified. We do not use that unresolved, grading-dependent identification as a completeness criterion: the explicit reduction and the agreement of the unrefined class-by-class counts with the direct massive-particle enumeration are the relevant checks. A systematic identification of the two refined gradings is left for future work.

\section{Type-II seesaw model and complete matching onto HEFT\label{sec:matching}}
As a concrete UV realization of LNV, we consider the type-II seesaw model and perform its tree-level matching onto HEFT through $\mathcal{O}(p^4)$. To the best of our knowledge, this is the first complete matching of this model within the HEFT framework, encompassing the bosonic and fermionic interactions in both the LNC and LNV sectors.

\subsection{The Type-II seesaw model}
\subsubsection{The Lagrangian}
The type-II seesaw model extends the SM by introducing a complex $SU(2)_L$ scalar triplet $\Xi$ with hypercharge $Y=1$~\cite{Konetschny:1977bn,Magg:1980ut,Cheng:1980qt,
Mohapatra:1980yp}. The part of the UV Lagrangian relevant to the
matching is
\begin{align}
\mathcal{L}^{c} \supset \left(D_{\mu}\mathcal{H}\right)^{\dagger}
                  \left(D^{\mu}\mathcal{H}\right)
                + \langle D_{\mu}\Xi^{\dagger} D^{\mu}\Xi \rangle
                - V^c_s - \mathcal{L}^c_Q - \mathcal{L}^c_L,
\label{eq:CHTM_Lag}
\end{align}
where the scalar potential and Yukawa interactions are given by
\begin{align}
V^c_s ={}&
  -m^{2}\mathcal{H}^{\dagger}\mathcal{H}
  + M^{2}\langle\Xi^{\dagger}\Xi\rangle
  + \bigl[\mu\,\mathcal{H}^{T}i\sigma_{2}\Xi^{\dagger}\mathcal{H}
          + \text{h.c.}\bigr]
  + \lambda_{1}(\mathcal{H}^{\dagger}\mathcal{H})^{2} \nonumber\\
&{} + \lambda_{2}\langle\Xi^{\dagger}\Xi\rangle^{2}
  + \lambda_{3}\langle\Xi^{\dagger}\Xi\Xi^{\dagger}\Xi\rangle
  + \lambda_{4}\mathcal{H}^{\dagger}\mathcal{H}\langle\Xi^{\dagger}\Xi\rangle
  + \lambda_{5}\mathcal{H}^{\dagger}\Xi\Xi^{\dagger}\mathcal{H},\\
  \mathcal{L}^c_Q ={}& y_u^{pr} \bar Q_L^p \widetilde{\mathcal H} u_R^r
+ y_d^{pr} \bar Q_L^p \mathcal{H} d_R^r
+ \text{h.c.}, \\
\mathcal{L}^c_L ={}& y_e^{pr} \bar L_L^p \mathcal{H} e_R^r
+ y_\Delta^{pr} L_L^{p T} C i\sigma_2 \Xi L_L^r
+ \text{h.c.}.
\label{eq:CHTM_V}
\end{align}
where $\widetilde{\mathcal H}\equiv i\sigma_2\mathcal H^*$, $p,r=1,2,3$ are flavor indices, and $C$ is the charge-conjugation matrix acting in spinor space. The triplet Yukawa matrix is symmetric in flavor space, $y_\Delta^{pr}=y_\Delta^{rp}$.

\subsubsection{The nonlinear $U$ representation}
Following Ref.~\cite{Song:2024kos}, we express the scalar doublet and triplet in the nonlinear $U$ representation as
\begin{align}
\mathcal{H} &= U \mathfrak{h},\quad 
 \quad U\equiv \exp\left(\frac{i \pi_i \sigma_i}{\vev}\right), \quad \mathfrak{h}\equiv
\begin{pmatrix}
\rho^+ \\
 \frac{1}{\sqrt{2}}(\vh+h^0+i \rho^0)
\end{pmatrix},
\label{eq:H} \\
\Xi&=U \boldsymbol{\Delta}U^\dagger,
    \quad \boldsymbol{\Delta}\equiv \frac{\Delta_a\sigma_a}{\sqrt{2}}=
\begin{pmatrix}
\Delta^+ & \Delta^{++} \\
 \Delta^0 & -\Delta^+
\end{pmatrix},
\label{eq:SigmaPhi}
\end{align}
where the matrix $U$ contains the three electroweak Goldstone bosons, whereas $\mathfrak{h}$ and $\Delta$ contain the remaining scalar degrees of freedom. The neutral triplet component is decomposed as $\Delta^{0}=(\vXi+\delta+i\eta)/\sqrt{2}$, where $v_\Xi$ is the triplet VEV. The electroweak scale is related to the doublet and triplet VEVs by $\vev=\sqrt{\vh^2 + 2 v_\Xi^2}$. The fields appearing in the doublet column are chosen according to
\begin{align}
    \rho^\pm = -2\,\frac{\vXi}{\vh}\,\Delta^\pm, 
    \qquad 
    \rho^0 = -2\,\frac{\vXi}{\vh}\,\eta,
\end{align}
so that the bilinear kinetic mixing between the Goldstone bosons and the remaining scalar fields vanishes. The Goldstone directions are therefore entirely contained in $U$ at the quadratic level, while the non-Goldstone scalar fields are collected only in $\mathfrak{h}$ and $\Delta$. This separation makes the subsequent integration of the heavy scalar fields particularly convenient for UV--HEFT matching.

\subsubsection{Physical states}
After electroweak symmetry breaking, the physical scalar spectrum consists of five types of mass eigenstates,
\begin{align}
    h, K, \eta, H^\pm, H^{\pm\pm}.
\end{align}
where $h$ is identified with the light, SM-like Higgs boson, $K$ is the additional CP-even neutral scalar, $\eta$ is the CP-odd neutral scalar, and $H^\pm$ and $H^{\pm\pm}$ are the singly and doubly charged scalars, respectively. The two CP-even fields $h^0$ and $\delta$ mix according to
\beq
\begin{pmatrix}
 h \\
 K\\
\end{pmatrix}
=
\begin{pmatrix}
 \cos \alpha & -\sin \alpha \\
 \sin \alpha & \cos \alpha
\end{pmatrix}
\begin{pmatrix}
 h^0 \\
 \delta \\
\end{pmatrix},
\label{eq:NSDiagMatrix}
\eeq
\beq
\tan 2\alpha =- \frac{\vXi}{\vh} \cdot \frac{2 \vh \lambda_{45} - \frac{2 \sqrt{2} \mu \vh}{\vXi}}{2 \vh \lambda_1 - \frac{\vh \mu}{\sqrt{2} \vXi} - \frac{2 \vXi^2 \lambda_{23}}{\vh}},
\eeq
where, for convenience, we define $\lambda_{23}\equiv\lambda_2+\lambda_3$ and $\lambda_{45}\equiv\lambda_4+\lambda_5$.

After imposing the stationarity conditions for the scalar potential, the Lagrangian parameters may be expressed in terms of the five physical scalar masses, the mixing angle $\alpha$, and the two VEVs as
\begin{align*}
\mu & = \frac{\sqrt{2} \vXi}{\vh^2 + 4 \vXi^2} m_\eta^2, \\
\lambda_1 &= \frac{1}{2 \vh^2} \left( m_h^2 \cos^2 \alpha + m_K^2 \sin^2 \alpha \right), \\
\lambda_2 &= \frac{1}{2 \vXi^2} \left[ 2 m_{++}^2 + \vh^2 \left( \frac{m_\eta^2}{\vh^2 + 4 \vXi^2} - \frac{4 m_{+}^2}{\vh^2 + 2 \vXi^2} \right) + m_K^2 \cos^2 \alpha + m_h^2 \sin^2 \alpha \right], \\
\lambda_3 &= \frac{\vh^2}{\vXi^2} \left( \frac{2 m_{+}^2}{\vh^2 + 2 \vXi^2} - \frac{m_{++}^2}{\vh^2} - \frac{m_\eta^2}{\vh^2 + 4 \vXi^2} \right), \\
\lambda_4 &= \frac{4 m_{+}^2}{\vh^2 + 2 \vXi^2} - \frac{2 m_\eta^2}{\vh^2 + 4 \vXi^2} + \frac{m_h^2 - m_K^2}{2 \vh \vXi} \sin 2\alpha, \\
\lambda_5 &= 4 \left( \frac{m_\eta^2}{\vh^2 + 4 \vXi^2} - \frac{m_{+}^2}{\vh^2 + 2 \vXi^2} \right),
\end{align*}
where $m_+$ and $m_{++}$ represent the masses of $H^\pm$ and $H^{\pm\pm}$, respectively. These relations are linear in the squared scalar masses. This property motivates the primary-HEFT parameterization adopted below, which preserves the dependence on the independent heavy-mass scales without introducing additional expansions or parameter relations~\cite{Ge:2026qfa}.

\subsection{Matching setup}
Following the primary-HEFT construction introduced in Ref.~\cite{Ge:2026qfa}, we adopt a parameterization in which the UV Lagrangian parameters depend linearly on the squared heavy masses. The resulting primary HEFT (pHEFT) serves as a benchmark description from which alternative HEFT power-counting schemes can be obtained by imposing additional scaling assumptions or parameter relations. We therefore perform the matching directly in the primary power-counting scheme.

We choose the input-parameter set
\beq
\{m_h,\; m_K,\; m_\eta,\; m_{+},\; m_{++},\; \sin\alpha,\; \vh,\; \xi\},
\eeq
which contains the five physical scalar masses, the CP-even mixing angle $\alpha$, the doublet VEV $\vh$, and the ratio of the triplet and doublet VEVs $\xi \equiv v_\Xi/\vh$. These quantities are taken as the independent parameters of the scalar sector after the stationarity conditions have been imposed.

We assign the scaling behavior
\beq
m^2_K,\; m^2_\eta,\; m^2_{+},\; m^2_{++}\;\sim \mathcal{O}(t^{-1}), \qquad 
\sin\alpha,\;\vh,\;\xi\;\sim \mathcal{O}(t^0),
\eeq
where $t$ is an auxiliary bookkeeping parameter. Equivalently, the inverse squared masses of the heavy scalar states scale as $\mathcal{O}(t)$. In the limit $t\ll 1$, the heavy scalar masses become parametrically large, while $\sin\alpha$, $\vh$, and $\xi$ remain finite quantities of $\mathcal{O}(t^0)$. The light-Higgs mass $m_h$ is also held fixed and therefore scales as $\mathcal{O}(t^0)$. No additional expansion in either the mixing angle or the VEV ratio is imposed.

\subsection{Tree-level matching}
At tree level, the heavy scalar fields are integrated out by solving their classical equations of motion (EoMs) and substituting the solutions back into the UV Lagrangian. The resulting effective Lagrangian contains only the light degrees of freedom, while the effects of the heavy fields are encoded in higher-order HEFT interactions. We solve the heavy-field EoMs perturbatively as a series in $t$. The resulting coefficient functions are then expanded in powers of $h$ to the order required for the interactions displayed below. The matching includes the independent HEFT operator structures generated through $\mathcal{O}(p^4)$ within the primary-HEFT power counting, while the Higgs-dependent coefficient functions are displayed to the indicated powers of $h$.

The matching results are reduced using integration-by-parts (IBP) relations and local field redefinitions of the light fields $h$ and $U$.\footnote{Since the pHEFT Lagrangian begins at $\mathcal{O}(t^{-1})$, a field redefinition of $h$ generates a quadratic contribution that is no longer suppressed relative to $\mathcal{O}(t)$, unlike in the conventional power counting; the EoM of $h$ is therefore no longer equivalent to a field redefinition, and the corresponding local field redefinitions are implemented explicitly instead. For $U$, the EoM reduction remains equivalent to a field redefinition because $\mathcal{L}_{\rm HEFT}^{(t^{-1})}$ is independent of $U$.} The reduced expressions are projected onto the complete minimal LNC basis of Ref.~\cite{Sun:2022ssa} and onto the LNV basis constructed in Sec.~\ref{sec:LNV}.

The reduced matching results are projected onto the minimal operator basis by expanding the fields and matrices appearing in each operator into explicit index components and solving the resulting linear system. This component-level representation automatically incorporates the relevant algebraic identities. For operators involving fermions, their anticommutation relations are implemented by treating the spinor components as Grassmann variables. 

We organize the tree-level matching results into three parts. We first present the bosonic contributions, including corrections to the Higgs potential, Higgs--Goldstone interactions, and higher-order derivative operators. We then discuss the fermionic operators in the LNC sector, including two-fermion interactions and four-fermion operators. Finally, we present the LNV contributions induced by the triplet Yukawa interactions, including the Majorana neutrino mass term and the $\Delta L=2$ and $\Delta L=4$ operators through $\mathcal O(p^4)$.

\subsubsection{Bosonic operators}
The HEFT Lagrangian is organized as an expansion in the auxiliary parameter $t$. The leading and next-to-leading contributions occur at $\mathcal{O}(t^{-1})$ and $\mathcal{O}(t^0)$, respectively. The corresponding bosonic terms are
\begin{align}
&\begin{aligned}[b]
\mathcal L_\text{HEFT}^{(t^{-1})}\supset{}&\frac{\vh^2\bigl[(4 \xi^2 + 1)m_K^2(\xi c + s)^2 - \xi^2 m_\eta^2\bigr]}{8(4 \xi^2 + 1)} - \frac{h^3 s^2 m_\eta^2(2 \xi c + s)}{2\xi(4 \xi^2 + 1)\vh} \\
& + \frac{h^4 s^2 m_\eta^2}{8 \xi^2(4 \xi^2 + 1)\vh^2}\biggl\{\frac{m_\eta^2(3 s c + 4 \xi - 6 \xi s^2)^2}{(4 \xi^2 + 1)m_K^2} \\
&\hspace{1.0cm} + 2 \xi c(9 s^2 - 7)s + 6 s^4 - 12 \xi^2(s^2 - 1)^2 - 5 s^2\biggr\} + \mathcal{O}(h^5) \\
\end{aligned} \\
&\begin{aligned}[b]
\mathcal L_\text{HEFT}^{(t^0)}\supset{}&\frac{1}{2}D_\mu h D^\mu h - \frac{1}{4}\braket{V_\mu V^\mu}\biggl\{(2 \xi^2 + 1)\vh^2 + 2 h \vh(c - 2 \xi s)+ h^2\biggl[1 + s^2 \\
&\hspace{1.0cm} - \frac{s(2 \xi c + s)\bigl(c(4 \xi^2 + 1)m_K^2(\xi c + s) - m_\eta^2(3 c s + 4 \xi - 6 \xi s^2)\bigr)}{\xi(4 \xi^2 + 1)m_K^2}\biggr] + \mathcal{O}(h^3)\biggr\} \\
& - \frac{1}{4} \braket{T V_\mu}\braket{T V^\mu}\biggl\{\xi^2 \vh^2 - 2 h \xi s \vh \\
&\hspace{1.0cm} - h^2 s\biggl[\xi c^3 - s^3 - \frac{c m_\eta^2(3 c s + 4 \xi - 6 \xi s^2)}{(4 \xi^2 + 1)m_K^2}\biggr] + \mathcal{O}(h^3)\biggr\} \\
& + \frac{1}{8}m_h^2 \vh^2(c - \xi s)^2 - \frac{1}{2}h^2 m_h^2 + \frac{h^3 m_h^2(s^3-\xi c^3)}{2 \xi \vh}+ \frac{h^4 m_h^2}{24 \xi^2(4 \xi^2 + 1)^2\vh^2 m_K^4} \\
&\hspace{1.0cm}\times \biggl\{(4 \xi^2 + 1)^2 m_K^4\bigl[c^4 \xi^2(19 s^2 - 3) + 38 c^3 \xi s^3 - 19 s^6 + 16 s^4\bigr] \\
&\hspace{2.0cm} - 20 c(4 \xi^2 + 1)s^2 m_K^2 m_\eta^2(c \xi + s)(3 c s + 4 \xi - 6 \xi s^2) \\
&\hspace{2.0cm} + 4 s^2 m_\eta^4(3 s c + 4 \xi - 6 \xi s^2)^2\biggr\} + \mathcal{O}(h^5) \\
\end{aligned}
\end{align}
where, hereafter, we use the abbreviations $s\equiv\sin\alpha$, $c\equiv\cos\alpha$. The expressions are displayed through the indicated powers of $h$, with the omitted terms denoted explicitly by $\mathcal{O}(h^n)$. At $\mathcal{O}(t^{-1})$, the bosonic Lagrangian contains only Higgs-potential terms. At $\mathcal{O}(t^0)$, two-derivative $\mathcal{O}(p^2)$ structures appear, including $D_\mu h D^\mu h$, $\braket{V_\mu V^\mu}$, and $\braket{TV_\mu}\braket{TV^\mu}$. At $\mathcal{O}(t)$, the matching generates both corrections to the lower-chiral-order coefficient functions and new $\mathcal{O}(p^4)$ operators. The relation between the $t$ expansion and the chiral order described here is specific to the primary power-counting scheme and should not be interpreted as a universal identification of the two expansions.

The bosonic contributions at $\mathcal O(t)$ are summarized in TABLEs~\ref{table:p2} and~\ref{table:p4}. For comparison, the third column of TABLE~\ref{table:p4} shows the corresponding results for the real-triplet model. The complex-triplet model contains the additional doubly charged scalar $H^{\pm\pm}$, which generates contributions that are not present in the real-triplet model. In particular, the operator
$\braket{V_\mu V_\nu}\braket{V^\mu V^\nu}$, shown in boldface in TABLE~\ref{table:p4}, has a Wilson coefficient proportional solely to $1/m_{H^{\pm\pm}}^2$, which is absent in the real-triplet model. The complex-triplet model also contains an additional CP-odd scalar $\eta$. Among the $\mathcal{O}(p^4)$ bosonic coefficients shown in TABLE~\ref{table:p4}, an explicit $\eta$-exchange contribution appears in the coefficient of $\braket{TV_\mu}\braket{TV_\nu}D^\mu h D^\nu h$ as indicated by the occurrence of $m_\eta^2$ in the denominator. For several other operators, including $\braket{V_\mu V^\mu} D_\nu h D^\nu h$, $\braket{T V_\mu}\braket{T V^\mu} D_\nu h D^\nu h$, and $D_\mu h D^\mu h D_\nu h D^\nu h$, the transition from the real- to the complex-triplet result instead amounts to replacing $m_{\phi^\pm}^2$ by $m_\eta^2$ in the numerator. Such terms should not be interpreted as additional $\eta$-exchange contributions; rather, they arise from the different parameterization of the scalar potential in the two models.
\begin{table}[h!]
  \centering
\begin{tabular}{cc}
    \hline
    \hline
   Operators & Wilson Coefficients (polynomials of h) \\
    \hline
    \hline
    $\braket{V_\mu V^\mu}$ &
    $\frac{h^2 s m_h^2(2 c \xi + s)\bigl[2 c(4 \xi^2 + 1)m_K^2(c \xi + s) - m_\eta^2(3 c s + 4 \xi - 6 \xi s^2)\bigr]}{2\xi(4 \xi^2 + 1)m_K^4}$ \\
    \hline
    $\braket{T V_\mu} \braket{T V^\mu}$ &
    $\frac{h^2 c s m_h^2\bigl[2 c(4 \xi^2 + 1)m_K^2(c \xi + s) - m_\eta^2(3 c s + 4 \xi - 6 \xi s^2)\bigr]}{2(4 \xi^2 + 1)m_K^4}$ \\
    \hline
    $V(h)$ &
    $\frac{h^4 s^2 m_h^4\bigl[8 c^2(4 \xi^2 + 1)^2 m_K^4(c \xi + s)^2 - 6 c(4 \xi^2 + 1)m_K^2 m_\eta^2(c \xi + s)(3 c s + 4 \xi - 6 \xi s^2) + m_\eta^4(3 c s + 4 \xi - 6 \xi s^2)^2\bigr]}{6 \xi^2(4 \xi^2 + 1)^2 \vh^2 m_K^6}$ \\
    \hline
\end{tabular}
\caption{Bosonic $\mathcal{O}(p^2)$ operators and pure Higgs-potential terms appearing in $\mathcal{L}_{\rm HEFT}^{(t^1)}$, together with their Wilson coefficients as polynomials in $h$.}
\label{table:p2}
\end{table}

\begin{table}[h!]
  \centering
\begin{tabular}{l|r|r}
    \hline
    \hline
   Operators & WC (Complex)& WC(Real) \\
    \hline
    \hline
    $\braket{V_\mu V^\mu} \braket{V_\nu V^\nu}$ &
    $\frac18\vh^2\Bigl[\frac{(2 \xi c + s)^2}{m_K^2} - \frac{4 \xi^2}{m_{++}^2}\Bigr]$ & $\frac{1}{8}\vh^2\frac{(4 \xi c_\gamma + s_\gamma)^2}{m_K^2}$\\
    \hline
    $\bm{\braket{V_\mu V_\nu} \braket{V^\mu V^\nu }}$ &
    $\frac{\xi^2 \vh^2}{m_{++}^2}$ &0\\
    \hline
    $\braket{T V_\mu} \braket{T V^\mu} \braket{V_\nu V^\nu}$ &
    $\frac14\xi \vh^2\Bigl[\frac{c (2 \xi c + s)}{m_K^2} + \frac{2 \xi}{m_{++}^2}\Bigr]$ & $ -\frac{1}{2} \xi \vh^2\frac{c_\gamma(4 \xi c_\gamma + s_\gamma)}{m_K^2}$\\
    \hline
    $\braket{T V_\mu} \braket{T V_\nu} \braket{V^\mu V^\nu}$ &
    $\frac12\xi^2 \vh^2\Bigl[\frac{1}{(2 \xi^2 + 1)m_{+}^2} - \frac{2}{m_{++}^2}\Bigr]$ &
    $\xi^2 \vh^2\frac{1}{(4 \xi^2 + 1)m_{\phi^\pm}^2}$\\
    \hline
    $\braket{T V_\mu} \braket{T V^\mu} \braket{T V_\nu} \braket{T V^\nu}$ &
    $\frac{1}{8}\xi^2 \vh^2\Bigl[\frac{c^2}{m_K^2} - \frac{2}{(2 \xi^2 + 1)m_{+}^2} + \frac{1}{m_{++}^2}\Bigr]$ &
    $\frac{1}{2}\xi^2 \vh^2\Bigl[\frac{c_\gamma^2}{m_K^2} - \frac{1}{(4 \xi^2 + 1)m_{\phi^\pm}^2}\Bigr]$\\
    \hline
    $\braket{T V_\mu V_\nu} \braket{T V^\mu} D^\nu h$ &
    $2\xi \vh\frac{\xi c + s}{(2 \xi^2 + 1)m_{+}^2}$ &
    $- 4 \xi \vh\frac{\xi c_\gamma + s_\gamma}{(4 \xi^2 + 1)m_{\phi^\pm}^2}$\\
    \hline
    $\braket{V_\mu V^\mu} D_\nu h D^\nu h$ &
    \makecell[r]{$- \tfrac{s(2 \xi c + s)}{2\xi(4 \xi^2 + 1)m_K^4}$ \\ $\times\Bigl[c(4 \xi^2 + 1)m_K^2(\xi c + s)$ \\ $- m_\eta^2(3 c s + 4 \xi - 6 \xi s^2)\Bigr]$} &
    \makecell[r]{$\tfrac{s_\gamma(4 \xi c_\gamma + s_\gamma)}{2\xi(4 \xi^2 + 1)m_K^4}$ \\ $\times\Bigl[m_{\phi^\pm}^2(3 c_\gamma s_\gamma + 4 \xi - 6 \xi s_\gamma^2)$ \\ $- (4 \xi^2 + 1)c_\gamma m_K^2(\xi c_\gamma + s_\gamma)\Bigr]$} \\
    \hline
    $\braket{T V_\mu} \braket{T V^\mu} D_\nu h D^\nu h$ &
    \makecell[r]{$- \tfrac{c s}{2(4 \xi^2 + 1)m_K^4}$ \\ $\times\Bigl[c(4 \xi^2 + 1)m_K^2(\xi c + s)$ \\ $- m_\eta^2(3 c s + 4 \xi - 6 \xi s^2)\Bigr]$} &
    \makecell[r]{$\tfrac{c_\gamma s_\gamma}{(4 \xi^2 + 1)m_K^4}$ \\ $\times\Bigl[(4 \xi^2 + 1)c_\gamma m_K^2(\xi c_\gamma + s_\gamma)$ \\ $- m_{\phi^\pm}^2(3 c_\gamma s_\gamma + 4 \xi - 6 \xi s_\gamma^2)\Bigr]$} \\
    \hline
    $\braket{V_\mu V_\nu} D^\mu h D^\nu h$ &
    $ - \frac{2(\xi c + s)^2}{(2 \xi^2 + 1)m_{+}^2}$ &
    $-\frac{4(\xi c_\gamma + s_\gamma)^2}{(4 \xi^2 + 1)m_{\phi^\pm}^2}$\\
    \hline
    $\braket{T V_\mu} \braket{T V_\nu} D^\mu h D^\nu h$ &
    $(\xi c + s)^2\Bigl[\frac{1}{(2 \xi^2 + 1)m_{+}^2} - \frac{2}{(4 \xi^2 + 1)m_\eta^2}\Bigr]$ & $2(\xi c_\gamma + s_\gamma)^2\frac{1}{(4 \xi^2 + 1)m_{\phi^\pm}^2}$\\
    \hline
    $D_\mu h D^\mu h D_\nu h D^\nu h$ &
    \makecell[r]{$\tfrac{s^2}{2 \xi^2(4 \xi^2 + 1)^2 \vh^2 m_K^6}$ \\ $\times\Bigl[c(4 \xi^2 + 1)m_K^2(\xi c + s)$ \\ $- m_\eta^2(3 c s + 4 \xi - 6 \xi s^2)\Bigr]^2$} &
    \makecell[r]{$\tfrac{s_\gamma^2}{2 \xi^2(4 \xi^2 + 1)^2 \vh^2 m_K^6}$ \\ $\times\Bigl[(4 \xi^2 + 1)c_\gamma m_K^2(\xi c_\gamma + s_\gamma)$ \\ $- m_{\phi^\pm}^2(3 c_\gamma s_\gamma + 4 \xi - 6 \xi s_\gamma^2)\Bigr]^2$} \\
    \hline
\end{tabular}
\caption{Bosonic $\mathcal{O}(p^4)$ operators in $\mathcal{L}_{\rm HEFT}^{(t^1)}$ and their Wilson coefficients. The second and third columns show the results for the complex triplet model (this work) and the real triplet model~\cite{Ge:2026qfa}, respectively. }
\label{table:p4}
\end{table}

\subsubsection{Fermionic operators in the LNC sector}
Operators involving two fermions first appear at $\mathcal{O}(t^0)$:
\beq
\mathcal{L}_{\rm HEFT}^{(t^0)} \supset 
- \frac{\vh + h c}{\sqrt{2}} \biggl[\bar Q_L U \begin{pmatrix}y_u&0\\0&y_d\end{pmatrix} Q_R + \bar L_L U \begin{pmatrix}0&0\\0&y_e\end{pmatrix} L_R + \text{h.c.}\biggr] + \mathcal{O}(h^2), 
\eeq
Four-fermion operators first arise at $\mathcal{O}(t)$. The two-fermion current operators are collected in TABLE~\ref{tab:fermion-current}, while the four-quark and the leptonic and semileptonic four-fermion operators are listed in TABLEs~\ref{tab:four-quark} and~\ref{tab:four-lepton}, respectively. The results retain the full flavor dependence, with $p,r,s,t$ denoting flavor indices, and are organized according to the complete HEFT basis of Ref.~\cite{Sun:2022ssa}.
\begin{table}[htbp]
  \centering
  \begin{tabular}{lr}
    \hline
    \hline
    Operators & Coefficients \\
    \hline
    \hline
    $\big(\bar Q_{Lp}UQ_{Rr}\big)$ &
    $\frac12(y_u^{pr}+y_d^{pr})\frac{h^2 s^2 m_h^2\bigl[2 c(4 \xi^2 + 1)m_K^2(c \xi + s) + m_\eta^2( - 3 c s - 4 \xi + 6 \xi s^2)\bigr]}{\sqrt{2}\xi(4 \xi^2 + 1)\vh m_K^4}$
    \\
    $\big(\bar Q_{Lp}TUQ_{Rr}\big)$ &
    $\frac12(y_u^{pr}-y_d^{pr})\frac{h^2 s^2 m_h^2\bigl[2 c(4 \xi^2 + 1)m_K^2(c \xi + s) + m_\eta^2( - 3 c s - 4 \xi + 6 \xi s^2)\bigr]}{\sqrt{2}\xi(4 \xi^2 + 1)\vh m_K^4}$
    \\
    \hline
    $\big(\bar Q_{Lp}UQ_{Rr}\big)\,D_\mu h D^\mu h$ &
    $-\frac12(y_u^{pr}+y_d^{pr})\frac{s^2\bigl[c(4 \xi^2 + 1)m_K^2(c \xi + s) + m_\eta^2( - 3 c s - 4 \xi + 6 \xi s^2)\bigr]}{\sqrt{2}\xi(4 \xi^2 + 1)\vh m_K^4}$
    \\
    $\big(\bar Q_{Lp}TUQ_{Rr}\big)\,D_\mu h D^\mu h$ &
    $-\frac12(y_u^{pr}-y_d^{pr})\frac{s^2\bigl[c(4 \xi^2 + 1)m_K^2(c \xi + s) + m_\eta^2( - 3 c s - 4 \xi + 6 \xi s^2)\bigr]}{\sqrt{2}\xi(4 \xi^2 + 1)\vh m_K^4}$
    \\
    \hline
    $\big(\bar Q_{Lp}V_\mu U Q_{Rr}\big)\,D^\mu h$ &
    $-2 \xi (y_d^{pr}+y_u^{pr})\frac{c \xi + s}{\sqrt{2}(2 \xi^2 + 1)m_+^2}$
    \\
    $\big(\bar Q_{Lp}[V_\mu,T]UQ_{Rr}\big)\,D^\mu h$ &
    $\xi (y_d^{pr}-y_u^{pr})\frac{c \xi + s}{\sqrt{2}(2 \xi^2 + 1)m_+^2}$
    \\
    $\big(\bar Q_{Lp}UQ_{Rr}\big)\braket{TV_\mu}D^\mu h$ &
    $\xi (y_d^{pr}-y_u^{pr})\frac{\sqrt{2}(c \xi + s)}{(4 \xi^2 + 1)m_\eta^2}$
    \\
    $\big(\bar Q_{Lp}TUQ_{Rr}\big)\braket{TV_\mu}D^\mu h$ &
    $-\xi (y_d^{pr}+y_u^{pr})\frac{\sqrt{2}(c \xi + s)}{(4 \xi^2 + 1)m_\eta^2}+\xi (y_d^{pr}+y_u^{pr})\frac{c \xi + s}{\sqrt{2}(2 \xi^2 + 1)m_+^2}$
    \\
    \hline
    $\big(\bar Q_{Lp}UQ_{Rr}\big)\braket{V_\mu V_\mu}$ &
    $\frac12(y_u^{pr}+y_d^{pr})\frac{s \vh(2 c \xi + s)}{2 \sqrt{2}m_K^2}$
    \\
    $\big(\bar Q_{Lp}TUQ_{Rr}\big)\braket{V_\mu V_\mu}$ &
    $\frac12(y_u^{pr}-y_d^{pr})\frac{s \vh(2 c \xi + s)}{2 \sqrt{2}m_K^2}$
    \\
    \hline
    $\big(\bar Q_{Lp}UQ_{Rr}\big)\braket{TV_\mu}\braket{TV^\mu}$ &
    $\frac12(y_u^{pr}+y_d^{pr})\frac{c \xi s \vh}{2 \sqrt{2}m_K^2}$
    \\
    $\big(\bar Q_{Lp}TUQ_{Rr}\big)\braket{TV_\mu}\braket{TV^\mu}$ &
    $\frac12(y_u^{pr}-y_d^{pr})\frac{c \xi s \vh}{2 \sqrt{2}m_K^2}+\xi (y_u^{pr}-y_d^{pr})\frac{\xi \vh}{2\sqrt{2}(2 \xi^2 + 1)m_+^2}$
    \\
    $\big(\bar Q_{Lp}[V_\mu,T]UQ_{Rr}\big)\braket{TV^\mu}$ &
    $-\xi (y_d^{pr}+y_u^{pr})\frac{\xi \vh}{2\sqrt{2}(2 \xi^2 + 1)m_+^2}$
    \\
    $\big(\bar Q_{Lp}V_\mu UQ_{Rr}\big)\braket{TV^\mu}$ &
    $-2 \xi (y_u^{pr}-y_d^{pr})\frac{\xi \vh}{2\sqrt{2}(2 \xi^2 + 1)m_+^2}$
    \\
    \hline
    $\big(\bar L_{Lp}UL_{Rr}\big)$ &
    $y_e^{pr}\frac{h^2 s^2 m_h^2\bigl[2 c(4 \xi^2 + 1)m_K^2(c \xi + s) + m_\eta^2( - 3 c s - 4 \xi + 6 \xi s^2)\bigr]}{\sqrt{2}\xi(4 \xi^2 + 1)\vh m_K^4}$
    \\
    \hline
    $\big(\bar L_{Lp}UL_{Rr}\big)\,D_\mu h D^\mu h$ &
    $-y_e^{pr}\frac{s^2\bigl[c(4 \xi^2 + 1)m_K^2(c \xi + s) + m_\eta^2( - 3 c s - 4 \xi + 6 \xi s^2)\bigr]}{\sqrt{2}\xi(4 \xi^2 + 1)\vh m_K^4}$
    \\
    \hline
    $\big(\bar L_{Lp}V_\mu U L_{Rr}\big)\,D^\mu h$ &
    $-4 \xi y_e^{pr}\frac{c \xi + s}{\sqrt{2}(2 \xi^2 + 1)m_+^2}$
    \\
    $\big(\bar L_{Lp}UL_{Rr}\big)\braket{TV_\mu}D^\mu h$ &
    $2 \xi y_e^{pr}\frac{\sqrt{2}(c \xi + s)}{(4 \xi^2 + 1)m_\eta^2}-2 \xi y_e^{pr}\frac{c \xi + s}{\sqrt{2}(2 \xi^2 + 1)m_+^2}$
    \\
    \hline
    $\big(\bar L_{Lp}UL_{Rr}\big)\braket{V_\mu V_\mu}$ &
    $y_e^{pr}\frac{s \vh(2 c \xi + s)}{2 \sqrt{2}m_K^2}$
    \\
    \hline
    $\big(\bar L_{Lp}UL_{Rr}\big)\braket{TV_\mu}\braket{TV^\mu}$ &
    $y_e^{pr}\frac{c \xi s \vh}{2 \sqrt{2}m_K^2}+2 \xi y_e^{pr}\frac{\xi \vh}{2\sqrt{2}(2 \xi^2 + 1)m_+^2}$
    \\
    $\big(\bar L_{Lp}V_\mu UL_{Rr}\big)\braket{TV^\mu}$ &
    $4 \xi y_e^{pr}\frac{\xi \vh}{2\sqrt{2}(2 \xi^2 + 1)m_+^2}$
    \\
    \hline
  \end{tabular}
  \caption{LNC two-fermion current operators and their corresponding Wilson coefficients at $\mathcal{O}(t)$.}
  \label{tab:fermion-current}
\end{table}

\begin{table}[h!]
  \centering
  \begin{tabular}{lr}
    \hline
    \hline
    Operators & Coefficients \\
    \hline
    \hline

    $\big(\bar Q_{Lp} U Q_{Rr}\big)\big(\bar Q_{Ls} U Q_{Rt}\big)$ &
    \makecell[r]{$\frac{1}{4} (y_d^{pr}+y_u^{pr}) (y_d^{st}+y_u^{st})\frac{s^2}{2 m_K^2}$ \\ $-\frac{1}{4} (y_d^{pr}-y_u^{pr}) (y_d^{st}-y_u^{st})\frac{2 \xi^2}{(4 \xi^2 + 1)m_\eta^2}$}
    \\
    $\big(\bar Q_{Lp} U Q_{Rr}\big)\big(\bar Q_{Ls} T U Q_{Rt}\big)$ &
    \makecell[r]{$-\frac{1}{2} (y_d^{pr}+y_u^{pr}) (y_d^{st}-y_u^{st})\frac{s^2}{2 m_K^2}$ \\
    $+\frac{1}{2} (y_d^{pr}-y_u^{pr}) (y_d^{st}+y_u^{st})\frac{2 \xi^2}{(4 \xi^2 + 1)m_\eta^2}$ \\
    $-y_u^{pr} y_d^{st}\frac{\xi^2}{2(2 \xi^2 + 1)m_+^2}$}
    \\
    $\big(\bar Q_{Lp} T U Q_{Rr}\big)\big(\bar Q_{Ls} T U Q_{Rt}\big)$ &
    \makecell[r]{$\frac{1}{4} (y_d^{pr}-y_u^{pr}) (y_d^{st}-y_u^{st})\frac{s^2}{2 m_K^2}$ \\ $-\frac{1}{4} (y_d^{pr}+y_u^{pr}) (y_d^{st}+y_u^{st})\frac{2 \xi^2}{(4 \xi^2 + 1)m_\eta^2}$ \\ $-y_d^{pr} y_u^{st}\frac{\xi^2}{2(2 \xi^2 + 1)m_+^2}$}
    \\
    $\big(\bar Q_{Lp} \sigma^I U Q_{Rr}\big)\big(\bar Q_{Ls} \sigma^I U Q_{Rt}\big)$ &
    $y_d^{pr} y_u^{st}\frac{\xi^2}{2(2 \xi^2 + 1)m_+^2}$
    \\
    $\big(\bar Q_{Lp} \sigma^I U Q_{Rr}\big)\big(\bar Q_{Ls} \sigma^I T U Q_{Rt}\big)$ &
    $y_d^{pr} y_u^{st}\frac{\xi^2}{2(2 \xi^2 + 1)m_+^2}$
    \\
    \hline
    $\big(\bar Q_{Lp}\gamma_\mu Q_{Lr}\big)\big(\bar Q_{Rs}\gamma^\mu Q_{Rt}\big)$ &
    $\frac{1}{4} (y_d^{pt} y_d^{rs*}+y_u^{pt} y_u^{rs*})\Bigl[\frac{s^2}{2 m_K^2}+\frac{2 \xi^2}{(4 \xi^2 + 1)m_\eta^2}-\frac{\xi^2}{(2 \xi^2 + 1)m_+^2}\Bigr]$
    \\
    $\big(\bar Q_{Lp}\gamma_\mu \sigma^I Q_{Lr}\big)\big(\bar Q_{Rs}\gamma^\mu U^\dagger \sigma^I U Q_{Rt}\big)$ &
    $\frac{1}{4} (y_u^{pt} y_d^{rs*}+y_d^{pt} y_u^{rs*})\Bigl[\frac{s^2}{2 m_K^2}-\frac{2 \xi^2}{(4 \xi^2 + 1)m_\eta^2}\Bigr]$
    \\
    $\big(\bar Q_{Lp}\gamma_\mu T Q_{Lr}\big)\big(\bar Q_{Rs}\gamma^\mu Q_{Rt}\big)$ &
    $\frac{1}{4} (y_u^{pt} y_u^{rs*}-y_d^{pt} y_d^{rs*})\Bigl[\frac{s^2}{2 m_K^2}+\frac{2 \xi^2}{(4 \xi^2 + 1)m_\eta^2}+\frac{\xi^2}{(2 \xi^2 + 1)m_+^2}\Bigr]$
    \\
    $\big(\bar Q_{Lp}\gamma_\mu \sigma^I T Q_{Lr}\big)\big(\bar Q_{Rs}\gamma^\mu U^\dagger \sigma^I U Q_{Rt}\big)$ &
    \makecell[r]{$-\frac{1}{8} (y_d^{pt}+y_u^{pt}) (y_d^{rs*}-y_u^{rs*})\frac{s^2}{2 m_K^2}$ \\ $-\frac{1}{8} (y_d^{pt}-y_u^{pt}) (y_d^{rs*}+y_u^{rs*})\frac{2 \xi^2}{(4 \xi^2 + 1)m_\eta^2}$ \\ $+\frac{1}{4} (y_d^{pt} y_d^{rs*}-y_u^{pt} y_u^{rs*})\frac{\xi^2}{2(2 \xi^2 + 1)m_+^2}$}
    \\
    $\big(\bar Q_{Lp}\gamma_\mu T Q_{Lr}\big)\big(\bar Q_{Rs}\gamma^\mu U^\dagger T U Q_{Rt}\big)$ &
    \makecell[r]{$\frac{1}{4} (y_d^{pt}-y_u^{pt}) (y_d^{rs*}-y_u^{rs*})\frac{s^2}{2 m_K^2}$ \\ $+\frac{1}{4} (y_d^{pt}+y_u^{pt}) (y_d^{rs*}+y_u^{rs*})\frac{2 \xi^2}{(4 \xi^2 + 1)m_\eta^2}$ \\ $+\frac{1}{2} (y_d^{pt} y_d^{rs*}+y_u^{pt} y_u^{rs*})\frac{\xi^2}{2(2 \xi^2 + 1)m_+^2}$}
    \\
    \hline
  \end{tabular}
  \caption{LNC four-quark operators and their corresponding Wilson coefficients at at $\mathcal{O}(t)$.}
  \label{tab:four-quark}
\end{table}

\begin{table}[h!]
  \centering
  \begin{tabular}{lr}
    \hline
    \hline
    Operators & Coefficients \\
    \hline
    \hline
     $\big(\bar L_{Lp}\gamma_\mu L_{Lr}\big)\big(\bar L_{Ls}\gamma^\mu L_{Lt}\big)$ &
    $-\frac{1}{4} y_\Delta^{rt} y_\Delta^{ps*}\Bigl[\frac{c^2}{4m_K^2} + \frac{1}{4(4 \xi^2 + 1)m_\eta^2} + \frac{1}{2m_{++}^2} + \frac{1}{(2 \xi^2 + 1)m_+^2}\Bigr]$
    \\
    $\big(\bar L_{Lp}\gamma_\mu L_{Lr}\big)\big(\bar L_{Ls}\gamma^\mu T L_{Lt}\big)$ &
    $-\frac{1}{8} \bigl(y_\Delta^{tr} y_\Delta^{ps*}+y_\Delta^{rt} y_\Delta^{sp*}\bigr)\Bigl[\frac{c^2}{2m_K^2} + \frac{1}{2(4 \xi^2 + 1)m_\eta^2} - \frac{1}{m_{++}^2}\Bigr]$
    \\
    $\big(\bar L_{Lp}\gamma_\mu T L_{Lr}\big)\big(\bar L_{Ls}\gamma^\mu T L_{Lt}\big)$ &
    $-\frac{1}{8} y_\Delta^{rt} y_\Delta^{ps*}\Bigl[\frac{c^2}{2m_K^2} + \frac{1}{2(4 \xi^2 + 1)m_\eta^2} + \frac{1}{m_{++}^2}\Bigr]+\frac{1}{2} y_\Delta^{tr} y_\Delta^{ps*}\frac{1}{2(2 \xi^2 + 1)m_+^2}$
    \\
    \hline
    $\big(\bar L_{Lp} U L_{Rr}\big)\big(\bar L_{Ls} U L_{Rt}\big)$ &
    $y_e^{pr} y_e^{st}\frac{s^2}{2 m_K^2}-y_e^{pr} y_e^{st}\frac{2 \xi^2}{(4 \xi^2 + 1)m_\eta^2}$
    \\
    \hline
    $\big(\bar L_{Lp}\gamma_\mu L_{Lr}\big)\big(\bar L_{Rs}\gamma^\mu L_{Rt}\big)$ &
    $\frac{1}{2} y_e^{pt} y_e^{rs*}\Bigl[\frac{s^2}{2 m_K^2}+\frac{2 \xi^2}{(4 \xi^2 + 1)m_\eta^2}-\frac{\xi^2}{(2 \xi^2 + 1)m_+^2}\Bigr]$
    \\
    $\big(\bar L_{Lp}\gamma_\mu \sigma^I L_{Lr}\big)\big(\bar L_{Rs}\gamma^\mu U^\dagger \sigma^I U L_{Rt}\big)$ &
    $\frac{1}{2} y_e^{pt} y_e^{rs*}\Bigl[\frac{s^2}{2 m_K^2}+\frac{2 \xi^2}{(4 \xi^2 + 1)m_\eta^2}+\frac{\xi^2}{(2 \xi^2 + 1)m_+^2}\Bigr]$
    \\
    \hline
    $\big(\bar L_{Lp} U L_{Rr}\big)\big(\bar Q_{Ls} U Q_{Rt}\big)$ &
    $y_e^{pr} (y_d^{st}+y_u^{st})\frac{s^2}{2 m_K^2}-y_e^{pr} (y_d^{st}-y_u^{st})\frac{2 \xi^2}{(4 \xi^2 + 1)m_\eta^2}$
    \\
    $\big(\bar L_{Lp} U L_{Rr}\big)\big(\bar Q_{Ls} T U Q_{Rt}\big)$ &
    \makecell[r]{$-y_e^{pr} (y_d^{st}-y_u^{st})\frac{s^2}{2 m_K^2}+y_e^{pr} (y_d^{st}+y_u^{st})\frac{2 \xi^2}{(4 \xi^2 + 1)m_\eta^2}$ \\ $+y_e^{pr} y_u^{st}\frac{\xi^2}{(2 \xi^2 + 1)m_+^2}$}
    \\
    $\big(\bar L_{Lp} \sigma^I U L_{Rr}\big)\big(\bar Q_{Ls} \sigma^I U Q_{Rt}\big)$ &
    $y_e^{pr} y_u^{st}\frac{\xi^2}{(2 \xi^2 + 1)m_+^2}$
    \\
    \hline
    $\big(\bar Q_{Lp}\gamma_\mu L_{Lr}\big)\big(\bar L_{Rs}\gamma^\mu Q_{Rt}\big)$ &
    $\frac{1}{2} y_d^{pt} y_e^{rs*}\Bigl[\frac{s^2}{2 m_K^2}+\frac{2 \xi^2}{(4 \xi^2 + 1)m_\eta^2}-\frac{\xi^2}{(2 \xi^2 + 1)m_+^2}\Bigr]$
    \\
    $\big(\bar Q_{Lp}\gamma_\mu T L_{Lr}\big)\big(\bar L_{Rs}\gamma^\mu Q_{Rt}\big)$ &
    \makecell[r]{$-\frac{1}{2} y_e^{rs*} (y_d^{pt}-y_u^{pt})\frac{s^2}{2 m_K^2}-\frac{1}{2} y_e^{rs*} (y_d^{pt}+y_u^{pt})\frac{2 \xi^2}{(4 \xi^2 + 1)m_\eta^2}$ \\ $-y_d^{pt} y_e^{rs*}\frac{\xi^2}{2(2 \xi^2 + 1)m_+^2}$}
    \\
    $\big(\bar Q_{Lp}\gamma_\mu \sigma^I L_{Lr}\big)\big(\bar L_{Rs}\gamma^\mu U^\dagger \sigma^I U Q_{Rt}\big)$ &
    $\frac{1}{2} y_u^{pt} y_e^{rs*}\Bigl[\frac{s^2}{2 m_K^2}-\frac{2 \xi^2}{(4 \xi^2 + 1)m_\eta^2}\Bigr]$
    \\
    \hline
  \end{tabular}
  \caption{LNC leptonic and semileptonic four-fermion operators and their Wilson coefficients at $\mathcal{O}(t)$.}
  \label{tab:four-lepton}
\end{table}

\subsubsection{Fermionic operators in the LNV sector}
The triplet Yukawa interaction and scalar mixing induce the leading Majorana-mass term and higher-order $\Delta L=2$ and $\Delta L=4$ interactions. Representative matching contributions are
\bea
\mathcal{L}_{\rm HEFT}^{(t^0)}  &\supset&-\frac{\xi\vh + h s}{\sqrt{2}} \biggl[L_L^T C i\sigma_2 U \begin{pmatrix}0&0\\y_\Delta&0\end{pmatrix} U^\dagger L_L + \text{h.c.}\biggr]~\label{eq:p2_Maj_mass}\\
\mathcal{L}_{\rm HEFT}^{(t^1)}&\supset& 
- \frac{\xi}{(4 \xi^2 + 1)m_\eta^2}
\biggl[L_L^T C i\sigma_2 U \begin{pmatrix}0&0\\y_\Delta&0\end{pmatrix} U^\dagger L_L \biggr]
\biggl[\bar Q_L U \begin{pmatrix}y_u&0\\0&y_d\end{pmatrix} Q_R \biggr]\nonumber\\
&& +\frac{1}{4}\biggl[\frac{c^2}{m_K^2} - \frac{1}{(4 \xi^2 + 1)m_\eta^2}\biggr]\biggl[L_L^T C i\sigma_2 U \begin{pmatrix}0&0\\y_\Delta&0\end{pmatrix} U^\dagger L_L\biggr]^2 +\text{h.c.}~\label{eq:p4_DL_exs}, 
\eea
The operator in Eq.~\eqref{eq:p2_Maj_mass} violates lepton number by two units, $\Delta L=2$, and contains both the Majorana neutrino mass term and the corresponding Higgs interaction. In Eq.~\eqref{eq:p4_DL_exs}, the semileptonic product in the first term likewise has $\Delta L=2$, whereas the squared leptonic bilinear in the second term carries $\Delta L=4$. These terms illustrate the $\Delta L=2$ and $\Delta L=4$ structures arising from the type-II seesaw. They are examples of operators in the general basis established in Sec.~\ref{sec:LNV}; the complete set of nonvanishing tree-level LNV matching coefficients is given below.

The nonvanishing LNV operators generated by the tree-level matching of the type-II seesaw model are listed in TABLE~\ref{tab:LNV_matching}. The $\Delta L=2$ and $\Delta L=4$ operator coefficients scale as $y_\Delta$ and $y_\Delta^2$, respectively. Additional factors of the SM Yukawa matrices appear in the coefficients of the semileptonic and mixed-chirality operators. The pattern of generated coefficients reflects the renormalizable interactions and scalar spectrum of the type-II seesaw model. Their phenomenological implications are discussed in Sec.~\ref{sec:pheno}. 
\begin{table}[htbp]
  \centering
  \begin{tabular}{lr}
    \hline
    \hline
    Operators & Coefficients \\
    \hline
    \hline
    $L_{Lp}^{T} C\widehat{\mathcal S} L_{Lr}$ &
    $y_\Delta^{pr}\frac{h^2 c s m_h^2\bigl[2 c(4 \xi^2 + 1)m_K^2(c \xi + s) + m_\eta^2( - 3 c s - 4 \xi + 6 \xi s^2)\bigr]}{\sqrt{2}\xi(4 \xi^2 + 1)\vh m_K^4}$
    \\
    \hline
    $L_{Lp}^{T} C\widehat{\mathcal S} L_{Lr}\,D_\mu h\,D^\mu h$ &
    $-y_\Delta^{pr}\frac{c s\bigl[c(4 \xi^2 + 1)m_K^2(c \xi + s) + m_\eta^2( - 3 c s - 4 \xi + 6 \xi s^2)\bigr]}{\sqrt{2}\xi(4 \xi^2 + 1)\vh m_K^4}$
    \\
    \hline
    $L_{Lp}^{T} C\widehat{\mathcal S}V_\mu L_{Lr}\,D^\mu h$ &
    $-2(y_\Delta^{pr}+y_\Delta^{rp})\frac{c \xi + s}{\sqrt{2}(2 \xi^2 + 1)m_+^2}$
    \\
    $L_{Lp}^{T} C\widehat{\mathcal S} L_{Lr}\braket{T V_\mu} D^\mu h$ &
    $-y_\Delta^{pr}\frac{\sqrt{2}(c \xi + s)}{(4 \xi^2 + 1)m_\eta^2}+2y_\Delta^{pr}\frac{c \xi + s}{\sqrt{2}(2 \xi^2 + 1)m_+^2}$
    \\
    \hline
    $L_{Lp}^{T} C\widehat{\mathcal S} L_{Lr}\braket{V_\mu V^\mu}$ &
    $y_\Delta^{pr}\frac{c \vh(2 c \xi + s)}{2 \sqrt{2}m_K^2}$
    \\
    $L_{Lp}^{T} C\widehat{\mathcal S} L_{Lr}\braket{T V_\mu}\braket{T V^\mu}$ &
    $y_\Delta^{pr}\frac{c^2 \xi \vh}{2 \sqrt{2}m_K^2}-2y_\Delta^{pr}\frac{\xi \vh}{2\sqrt{2}(2 \xi^2 + 1)m_+^2}$
    \\
    $L_{Lp}^{T} C \widehat{\mathcal S}V_\mu L_{Lr}\braket{T V^\mu}$ &
    $2(y_\Delta^{pr}+y_\Delta^{rp})\frac{\xi \vh}{2\sqrt{2}(2 \xi^2 + 1)m_+^2}$
    \\
    $L_{Lp}^{T} C\epsilon\widehat{\mathcal S}^{\dagger}\epsilon L_{Lr}\braket{\epsilon\widehat{\mathcal S}V_\mu}\braket{\epsilon\widehat{\mathcal S}V^\mu}$ &
    $-4y_\Delta^{pr}\frac{\xi \vh}{2 \sqrt{2}m_{++}^2}$
    \\
    \hline
    $\displaystyle \big(L_{Lp}^{T}C\widehat{\mathcal S}L_{Lr}\big)\big(L_{Ls}^{T}C\widehat{\mathcal S}L_{Lt}\big)$ &
    $y_\Delta^{pr}y_\Delta^{st}\frac{1}{4}\Bigl[\frac{c^2}{m_K^2} - \frac{1}{(4 \xi^2 + 1)m_\eta^2}\Bigr]$
    \\
    \hline
    $\big(L_{Lp}^{T}C\widehat{\mathcal S}L_{Lr}\big)\big(\bar L_{Ls}U L_{Rt}\big)$ &
    $y_\Delta^{pr} y_e^{st}\frac{c s}{2 m_K^2}+y_\Delta^{pr} y_e^{st}\frac{\xi}{(4 \xi^2 + 1)m_\eta^2}$
    \\
    $\big(L_{Lp}^{T}C\widehat{\mathcal S}L_{Lr}\big)\big(\bar L_{Rs}U^\dagger L_{Lt}\big)$ &
    \makecell[r]{$-y_\Delta^{pr} y_e^{ts*}\frac{c s}{2 m_K^2}+y_\Delta^{pr} y_e^{ts*}\frac{\xi}{(4 \xi^2 + 1)m_\eta^2}$ \\ $+\bigl[y_\Delta^{pr} y_e^{ts*}-\bigl(y_\Delta^{pt}+y_\Delta^{tp}\bigr) y_e^{rs*}\bigr]\frac{\xi}{(2 \xi^2 + 1)m_+^2}$}
    \\
    $\big(L_{Lp}^{T}C\epsilon L_{Lr}\big)\big(\bar L_{Rs}U^\dagger\epsilon\widehat{\mathcal S}L_{Lt}\big)$ &
    $-(y_\Delta^{pt}+y_\Delta^{tp}) y_e^{rs*}\frac{\xi}{(2 \xi^2 + 1)m_+^2}$
    \\
    \hline
    $\big(L_{Lp}^{T}C\widehat{\mathcal S}L_{Lr}\big)\big(\bar Q_{Ls}UQ_{Rt}\big)$ &
    $\frac{1}{2} y_\Delta^{pr} (y_d^{st}+y_u^{st})\frac{c s}{2 m_K^2}+\frac{1}{2} y_\Delta^{pr} (y_d^{st}-y_u^{st})\frac{\xi}{(4 \xi^2 + 1)m_\eta^2}$
    \\
    $\big(L_{Lp}^{T}C\widehat{\mathcal S}L_{Lr}\big)\big(\bar Q_{Ls}TUQ_{Rt}\big)$ &
    $-\frac{1}{2} y_\Delta^{pr} (y_d^{st}-y_u^{st})\frac{c s}{2 m_K^2}-\frac{1}{2} y_\Delta^{pr} (y_d^{st}+y_u^{st})\frac{\xi}{(4 \xi^2 + 1)m_\eta^2}$
    \\
    $\big(L_{Lp}^{T}C\widehat{\mathcal S}L_{Lr}\big)\big(\bar Q_{Rs}U^\dagger Q_{Lt}\big)$ &
    $-\frac{1}{2} y_\Delta^{pr} (y_d^{ts*}+y_u^{ts*})\frac{c s}{2 m_K^2}+\frac{1}{2} y_\Delta^{pr} (y_d^{ts*}-y_u^{ts*})\frac{\xi}{(4 \xi^2 + 1)m_\eta^2}$
    \\
    $\big(L_{Lp}^{T}C\widehat{\mathcal S}L_{Lr}\big)\big(\bar Q_{Rs}U^\dagger TQ_{Lt}\big)$ &
    $\frac{1}{2} y_\Delta^{pr} (y_d^{ts*}-y_u^{ts*})\frac{c s}{2 m_K^2}-\frac{1}{2} y_\Delta^{pr} (y_d^{ts*}+y_u^{ts*})\frac{\xi}{(4 \xi^2 + 1)m_\eta^2}$
    \\
    $\big(L_{Lp}^{T}C\epsilon T L_{Lr}\big)\big(\bar Q_{Ls}\epsilon\widehat{\mathcal S}UQ_{Rt}\big)$ &
    $y_\Delta^{pr} y_u^{st}\frac{\xi}{(2 \xi^2 + 1)m_+^2}$
    \\
    $\big(L_{Lp}^{T}C\epsilon T L_{Lr}\big)\big(\bar Q_{Rs}U^\dagger\epsilon\widehat{\mathcal S}Q_{Lt}\big)$ &
    $y_\Delta^{pr} y_d^{ts*}\frac{\xi}{(2 \xi^2 + 1)m_+^2}$
    \\
    \hline
  \end{tabular}
  \caption{Nonvanishing LNV HEFT operators generated by the tree-level matching of the type-II seesaw model and their corresponding Wilson coefficients.}
\label{tab:LNV_matching}
\end{table}

\subsection{Comparison with the broken-phase effective description}\label{sec:bEFT_comparison}
It is instructive to compare our matching results with the broken-phase effective field theory (bEFT) developed in Ref.~\cite{Liao:2025bmn}, in which the heavy physical scalars are integrated out after electroweak symmetry breaking. That study also considered Higgs-pair production through vector-boson fusion and single-Higgs production through Higgsstrahlung. In contrast, our HEFT matching organizes the resulting interactions into operators respecting the nonlinearly realized electroweak symmetry, with general Higgs-dependent coefficient functions. To compare the two descriptions, we work in the unitary gauge, translate between the respective parameter conventions, and expand our results in $\epsilon\equiv v_\Xi/v_{\rm EW}$. We find that the corresponding bEFT Wilson coefficients are reproduced in the overlapping expansion, providing a nontrivial check of our matching results. The explicit coefficient-by-coefficient comparisons for operators of canonical dimensions three and four are presented in Tables~\ref{tab:bEFT_comparison} and \ref{tab2:bEFT_comparison} of Appendix~\ref{app:bEFT_details}. We have also verified the agreement for the corresponding dimension-five and dimension-six operators. Before this expansion, our HEFT coefficients retain their unexpanded dependence on the physical heavy-scalar masses, the neutral-scalar mixing angle, and the triplet VEV, including terms beyond the small-$\epsilon$ expansion used in the comparison. A quantitative comparison of the two effective descriptions for specific collider observables would require a dedicated comparison with the corresponding full type-II seesaw amplitudes and cross sections, which is beyond the scope of the present work.

\section{Phenomenology\label{sec:pheno}}
The HEFT operators obtained above contribute to several low- and high-energy observables. We first discuss the LNV interactions relevant to neutrinoless double-beta decay and same-sign dilepton production. We then consider charged-lepton flavor violation, Higgs and electroweak precision observables, vector-boson scattering, multi-Higgs production, and top-quark processes. In particular, $\mu\to3e$ conserves total lepton number, although it probes the same triplet Yukawa couplings that enter the LNV sector.

\subsection{Lepton-number-violating probes and neutrinoless double-beta decay}
The LNV operator basis constructed in Sec.~\ref{sec:LNV} provides a systematic HEFT description of low-energy lepton-number-violating observables. At the level of the general EFT, the coefficients of these operators need not obey the neutrino-mass and heavy-scalar relations that follow from the type-II seesaw matching. The complete basis contains both $\Delta L=2$ and $\Delta L=4$ interactions, whereas neutrinoless double-beta decay ($0\nu\beta\beta$) is a $\Delta L=2$ process and directly probes the former sector. At leading chiral order, the operator $\mathcal{O}_L^{pr}=L_{Lp}^{T}C\hat{\mathcal{S}}L_{Lr}$ contains the Majorana neutrino mass term and therefore generates the standard light-neutrino-exchange contribution to $0\nu\beta\beta$. Beyond this leading contribution, the $\Delta L=2$ basis contains operators involving $\hat{\mathcal S}$ together with Goldstone currents, electroweak field strengths, and fermion currents. After electroweak symmetry breaking, these operators generate nonstandard charged-current interactions that can contribute to $0\nu\beta\beta$.

HEFT therefore provides an electroweak-scale description in which the nonlinear realization of electroweak symmetry and the custodial structure of the LNV interactions remain manifest. The relevant operators can subsequently be matched onto LEFT and evolved toward the hadronic and nuclear scales. In Appendix~\ref{app:0vbb_matching}, we provide the HEFT-to-LEFT matching relations for the operators relevant at the chiral orders considered here. These relations translate the HEFT Wilson coefficients into LEFT coefficients that can be used in EFT-based calculations of $0\nu\beta\beta$ observables, including those implemented in $\nu$DoBe~\cite{Scholer:2023bnn}. Since our analysis is truncated at $\mathcal{O}(p^4)$, it captures the Majorana-mass contribution and the associated lower-order nonstandard interactions, whereas the complete short-range sector lies beyond the present operator set.

\subsection{Same-sign dilepton probes of lepton-number violation}
Same-sign dilepton production provides a high-energy probe of the $\Delta L=2$ interactions discussed above and is complementary to low-energy observables such as neutrinoless double-beta decay. A representative hadron-collider signature is $pp\to\ell_p^\pm\ell_r^\pm jj$ with the underlying electroweak subprocess $W^\pm W^\pm\to\ell_p^\pm\ell_r^\pm$. In a general HEFT, this amplitude receives two qualitatively different contributions. The leading Majorana operator $\mathcal O_L^{pr}=L_{Lp}^{T}C\hat{\mathcal S}L_{Lr}$ generates the light-neutrino Majorana mass and hence contributes through light-Majorana-neutrino exchange between two charged-current vertices. In addition, the $L^2 V^2$ sector contains local interactions involving two same-sign weak bosons and two charged leptons. The operators that directly generate such charged contact interactions are $\mathcal{O}_{LVV,4}^{pr}=\big(L_{Lp}^{T}C\epsilon\hat{\mathcal S}^{\dagger}\epsilon L_{Lr}\big)\braket{\epsilon\hat{\mathcal S}V_\mu}\braket{\epsilon\hat{\mathcal S}V^\mu}$ and $\mathcal{O}_{LVV,7}^{pr}=\big(L_{Rp}^{T}C U^T\epsilon\widehat{\mathcal S}^{\dagger}\epsilon U L_{Rr}\big)\braket{\epsilon\widehat{\mathcal S}V_\mu}\braket{\epsilon\widehat{\mathcal S}V^\mu}$. In the unitary gauge, their charged components take the schematic form $\mathcal O_{LVV,4}^{pr}\supset W_\mu^+W^{+\mu}\overline{e_{Lp}^{c}}e_{Lr}$, $\mathcal O_{LVV,7}^{pr}\supset W_\mu^+W^{+\mu}\overline{e_{Rp}^{c}}e_{Rr}$ together with their Hermitian conjugates and up to the normalization associated with $V_\mu$. These operators therefore describe, respectively, left- and right-handed contact contributions to $W^\pm W^\pm\to\ell_p^\pm\ell_r^\pm$. The remaining $L^2 V^2$ operators generate electroweakly related $\Delta L=2$ interactions involving neutrinos or neutral gauge bosons, but do not directly contain this charged contact vertex.

For the tree-level type-II seesaw matching, only the left-handed operator is generated. Denoting its coefficient at $h=0$ by $C_{LVV,4}^{pr}$, we obtain
\begin{align}
    C_{LVV,4}^{pr}=-\frac{\sqrt{2}\,y_\Delta^{pr}v_\Xi}{m_{H^{\pm\pm}}^2},\qquad C_{LVV,7}^{pr}=0.
    \label{eq:typeII_same_sign}
\end{align}
This contact interaction is the low-energy limit of off-shell $H^{\pm\pm}$ exchange. The coupling of $H^{\pm\pm}$ to two same-sign weak bosons is controlled by the triplet VEV $v_\Xi$, whereas its coupling to charged leptons is controlled by $y_\Delta^{pr}$. Consequently, the nonresonant process is sensitive to the combination $y_\Delta^{pr}v_\Xi/m_{H^{\pm\pm}}^2$. Since the same product $y_\Delta^{pr}v_\Xi$ determines the Majorana neutrino mass matrix, the type-II matching correlates the leading Majorana operator with the higher-order $L^2 V^2$ contact interaction.

Nonresonant same-sign dilepton production through weak-boson fusion has been studied as an EFT probe of the Weinberg operator and has also been tested experimentally at the LHC~\cite{Fuks:2020zbm,CMS:2022fut}. The type-II scalar sector has meanwhile been investigated directly through on-shell production of $H^{\pm\pm}$ followed by its decay into same-sign leptons, including pair-production, associated-production, and vector-boson-fusion channels~\cite{Dev:2018sel,Fuks:2019clu,ATLAS:2022pbd}. Pair-production searches primarily constrain the doubly charged scalar mass and its leptonic branching fractions, whereas $W^\pm W^\pm$-fusion production followed by $H^{\pm\pm}\to\ell_p^\pm\ell_r^\pm$ probes both $v_\Xi$ and $y_\Delta^{pr}$ and is the resonant counterpart of the HEFT contact interaction. The two descriptions therefore apply in complementary kinematic regimes: the full type-II seesaw model is required near an $H^{\pm\pm}$ resonance, while HEFT systematically organizes its virtual effects at energies below the heavy-scalar scale.

The relation $y_\Delta^{pr}v_\Xi\propto(m_\nu)^{pr}$ further clarifies the interpretation of the local contact term in Eq.~\eqref{eq:typeII_same_sign}. In the heavy-triplet regime where this HEFT description applies, its coefficient is proportional to a neutrino-mass-matrix entry divided by $m_{H^{\pm\pm}}^2$. The resulting nonresonant amplitude is therefore strongly suppressed in the minimal type-II seesaw. The correlation between the leading Majorana interaction and the $L^2V^2$ contact coefficient is a definite result of the matching, although the corresponding nonresonant signal is strongly suppressed in this model.

\subsection{Charged-lepton flavor violation in $\mu\to 3e$}
The decay $\mu\to 3e$ provides a sensitive probe of both the four-lepton sector of HEFT and, through the matching, the underlying flavor structure of the type-II seesaw model. Although it violates charged-lepton flavor, the process conserves total lepton number and must not be interpreted as a direct observation of $\Delta L=2$ interactions. Its role here is to illustrate how the same UV triplet couplings that enter the LNV sector also constrain LNC flavor-changing observables. The relevant HEFT interactions can be written compactly as
\begin{align}
    \mathcal L_{\mu\to 3e}^{\rm HEFT}={}&\mathcal C_{1}^{prst}(\bar L_{Lp}\gamma_\mu L_{Lr})(\bar L_{Ls}\gamma^\mu L_{Lt})+\mathcal C_{2}^{prst}(\bar L_{Lp}\gamma_\mu L_{Lr})(\bar L_{Ls}\gamma^\mu T L_{Lt}) \nonumber \\
    &+\mathcal C_{3}^{prst}(\bar L_{Lp}\gamma_\mu T L_{Lr})(\bar L_{Ls}\gamma^\mu T L_{Lt})+\left[\mathcal C_{4}^{prst}(\bar L_{Lp}U L_{Rr})(\bar L_{Ls}U L_{Rt})+\text{h.c.}\right] \nonumber \\
    &+\mathcal C_{5}^{prst}(\bar L_{Lp}\gamma_\mu L_{Lr})(\bar L_{Rs}\gamma^\mu L_{Rt})+\mathcal C_{6}^{prst}(\bar L_{Lp}\gamma_\mu\sigma^I L_{Lr})(\bar L_{Rs}\gamma^\mu U^\dagger\sigma^I U L_{Rt}) \nonumber \\
    &+\mathcal C_{7}^{prst}(\bar L_{Rp}\gamma_\mu L_{Rr})(\bar L_{Rs}\gamma^\mu L_{Rt})~\label{eq:HEFT_mu3e}
\end{align}
After electroweak symmetry breaking and projection onto the charged-lepton components, they match onto the LEFT Lagrangian~\cite{Jenkins:2017jig}
\begin{align}
    \mathcal L_{\mu\to 3e}^{\rm LEFT}={}&C_{VLL}(\bar e_L\gamma_\mu\mu_L)(\bar e_L\gamma^\mu e_L)+C_{VRR}(\bar e_R\gamma_\mu\mu_R)(\bar e_R\gamma^\mu e_R) \nonumber\\
    &+C_{VLR}(\bar e_L\gamma_\mu\mu_L)(\bar e_R\gamma^\mu e_R)+C_{VRL}(\bar e_R\gamma_\mu\mu_R)(\bar e_L\gamma^\mu e_L) \nonumber\\
    &+C_{SRR}(\bar e_L\mu_R)(\bar e_L e_R)+C_{SLL}(\bar e_R\mu_L)(\bar e_R e_L)+{\rm h.c.}~\label{eq:LEFT_mu3e}
\end{align}
with
\begin{equation}
    \begin{gathered}
        C_{VLL}=\mathcal C_1^{e\mu ee}+\mathcal C_1^{eee\mu}-\mathcal C_2^{e\mu ee}-\mathcal C_2^{eee\mu}+\mathcal C_3^{e\mu ee}+\mathcal C_3^{eee\mu}, \nonumber\\
        C_{VRR}=\mathcal C_7^{e\mu ee}+\mathcal C_7^{eee\mu},\qquad C_{VLR}=\mathcal C_5^{e\mu ee}+\mathcal C_6^{e\mu ee},\qquad C_{VRL}=\mathcal C_5^{eee\mu}+\mathcal C_6^{eee\mu}, \nonumber\\
        C_{SRR}=\mathcal C_4^{e\mu ee}+\mathcal C_4^{eee\mu},\qquad C_{SLL}=\left(\mathcal C_4^{\mu eee}+\mathcal C_4^{ee\mu e}\right)^* .~\label{eq:HEFT_LEFT_mu3e}
    \end{gathered}
\end{equation}
Neglecting the electron mass, the branching fraction is
\begin{align}
    {\rm BR}(\mu\to3e)={}&\frac{|C_{VLL}|^2+|C_{VRR}|^2}{4G_F^2}+\frac{|C_{VLR}|^2+|C_{VRL}|^2}{8G_F^2}+\frac{|C_{SRR}|^2+|C_{SLL}|^2}{64G_F^2}.~\label{eq:BR_mu3e_LEFT}
\end{align}

For the tree-level type-II seesaw matching, the charged-lepton Yukawa matrix is diagonal in the charged-lepton mass basis, and only the left-handed vector coefficient contributes. In the conventions adopted here, the result is
\begin{gather}
    C_{VLL}^{\rm II}=-\frac{y_\Delta^{e\mu}y_\Delta^{ee*}}{m_{H^{\pm\pm}}^2}, \\
    C_{VRR}=C_{VLR}=C_{VRL}=C_{SRR}=C_{SLL}=0,
\end{gather}
and therefore
\begin{align}
    {\rm BR}(\mu\to3e)_{\rm II}=\frac{|y_\Delta^{e\mu}y_\Delta^{ee*}|^2}{4G_F^2m_{H^{\pm\pm}}^4}.~\label{eq:BR_mu3e_typeII}
\end{align}
The decay therefore constrains the magnitude of the flavor combination $y_\Delta^{e\mu}y_\Delta^{ee*}/m_{H^{\pm\pm}}^2$. Upon imposing the type-II seesaw relation $m_\nu\propto y_\Delta v_\Xi$, this bound can also be translated into a constraint on the corresponding entries of the Majorana neutrino mass matrix. The process has been studied directly in the full type-II seesaw model in both dedicated analyses of charged-lepton flavor violation~\cite{Akeroyd:2009nu,Dinh:2012bp,Barrie:2022ake} and broader studies that incorporate $\mu\to3e$ among the constraints on the viable model parameter space~\cite{Dev:2018sel,Ashanujjaman:2025scr}. The current SINDRUM limit, ${\rm BR}(\mu\to3e)<1.0\times10^{-12}$ at $90\%$ confidence level, already imposes a stringent constraint, while Mu3e is designed to reach a single-event sensitivity of approximately $2\times10^{-15}$ in Phase~I and ultimately probe branching fractions at the $10^{-16}$ level~\cite{SINDRUM:1987nra,Mu3e:2020gyw}. Our analysis reproduces the direct full-model result through the HEFT-to-LEFT matching and, more generally, identifies the combination of four-lepton HEFT coefficients tested by this decay.

\subsection{Higgs and electroweak precision observables}
The leading bosonic terms obtained from the matching directly encode the effects of the scalar triplet on electroweak precision observables and Higgs couplings. The gauge-boson masses are
\begin{align}
    m_W^2&=\frac{g^2}{4}\left(\vh^2+2v_\Xi^2\right)=\frac{g^2v_{\rm EW}^2}{4}, \\
    m_Z^2&=\frac{g^2+g'^2}{4}\left(\vh^2+4v_\Xi^2\right).
\end{align}
Consequently, the tree-level rho parameter is
\begin{align}
    \rho_{\rm tree}\equiv\frac{m_W^2}{m_Z^2c_W^2}=\frac{\vh^2+2v_\Xi^2}{\vh^2+4v_\Xi^2}=\frac{1+2\xi^2}{1+4\xi^2}=1-2\xi^2+\mathcal O(\xi^4).~\label{eq:rho_tree}
\end{align}
The deviation of $\rho_{\rm tree}$ from unity reflects the custodial-symmetry-breaking contribution of the triplet VEV and therefore places a strong constraint on $\xi=v_\Xi/\vh$.

The same matched terms determine the couplings of the physical Higgs boson to the electroweak gauge bosons. Normalizing them to their SM values, we obtain
\begin{align}
    \kappa_W&\equiv\frac{g_{hWW}}{g_{hWW}^{\rm SM}}=\frac{c-2\xi s}{\sqrt{1+2\xi^2}}=c-2\xi s-c\xi^2+\mathcal{O}(\xi^3), \\
    \kappa_Z&\equiv\frac{g_{hZZ}}{g_{hZZ}^{\rm SM}}=\frac{c-4\xi s}{\sqrt{1+2\xi^2}}=c-4\xi s-c\xi^2+\mathcal{O}(\xi^3),~\label{eq:kappa_WZ}
\end{align}
The difference between $\kappa_W$ and $\kappa_Z$ provides a direct Higgs-sector probe of custodial-symmetry breaking. The fermionic Higgs couplings are modified according to
\begin{align}
    \kappa_f\equiv\frac{g_{hff}}{g_{hff}^{\rm SM}}=c\sqrt{1+2\xi^2}=c+c\xi^2+\mathcal{O}(\xi^4).
\end{align}

The expressions above already imply tree-level correlations among quantities that are independent parameters in a general HEFT. For example, in the normalization adopted here,
\begin{align}
  \kappa_W-\kappa_Z&=\frac{2\xi s}{\sqrt{1+2\xi^2}}, &
  2\kappa_W-\kappa_Z&=\frac{\kappa_f}{1+2\xi^2}.\label{eq:kappa_matching_correlations}
\end{align}
Eliminating $\xi$ by means of Eq.~\eqref{eq:rho_tree} gives an equivalent relation directly among the displayed observables,
\begin{align}
    \rho_{\rm tree}(2\kappa_W-\kappa_Z)=(2\rho_{\rm tree}-1)\kappa_f.\label{eq:kappa_rho_relation}
\end{align}
These relations display how the type-II seesaw embeds a restricted pattern of Higgs and electroweak couplings into the more general HEFT parameter space. They follow algebraically from the tree-level matching already presented, in the normalizations adopted here, and do not constitute a separate estimate of experimental sensitivity or model discrimination.

The deviations of electroweak precision observables and Higgs couplings from their SM values have been extensively constrained in model-independent analyses based on HEFT, using electroweak precision data, Higgs measurements, and diboson production~\cite{Brivio:2016fzo,deBlas:2018tjm,Eboli:2021unw}. These analyses constrain the corresponding low-energy HEFT coefficients rather than the parameters of a particular UV model. Through the matching relations derived above, their bounds can be translated into constraints on the triplet VEV ratio $\xi$ and the doublet--triplet mixing angle $\alpha$. In particular, $\rho_{\rm tree}$ primarily constrains $\xi$, whereas Higgs-coupling measurements probe the combinations $c-2\xi s$, $c-4\xi s$, and $c$. A dedicated global interpretation in terms of the type-II seesaw parameters is beyond the scope of the present work. It would nevertheless be informative to compare the constraints inferred from the HEFT matching with those obtained from direct analyses of the full type-II seesaw model, including electroweak precision fits and studies of Higgs couplings, decay observables, and direct scalar searches~\cite{Chen:2006pb,Aoki:2012jj,Ashanujjaman:2025scr,Aiko:2026qzi}~\footnote{We note that some of these studies include loop-level effects that lie beyond the scope of our tree-level matching. Several of them also consider additional observables beyond Higgs and electroweak precision measurements, some of which are related to the phenomenological applications discussed below.}.

\subsection{Vector-boson scattering, multi-Higgs production, and anomalous gauge couplings}
Vector-boson scattering (VBS) and multi-Higgs production provide sensitive probes of the electroweak symmetry-breaking sector. In HEFT, the electroweak Goldstone bosons are realized nonlinearly, whereas the physical Higgs field is treated as an independent singlet. In a general HEFT, interactions involving longitudinal gauge bosons and different numbers of Higgs bosons are therefore controlled by independent coefficients in the expansions of the Higgs-dependent functions. SMEFT, by contrast, occupies a restricted subspace of the general HEFT parameter space and imposes additional correlations among interactions involving different Higgs multiplicities~\cite{Gomez-Ambrosio:2022qsi,Gomez-Ambrosio:2022why}. For a given UV completion, such as the type-II seesaw model considered here, these otherwise independent HEFT coefficients become correlated through the matching conditions.

These differences can be tested through longitudinal VBS and vector-boson-fusion production of multiple Higgs bosons. Processes such as $V_L V_L\to V_L V_L$, $V_L V_L\to V_L V_L h$, $V_L V_L\to hh$, and $V_L V_L\to n h$ probe complementary combinations of gauge--Higgs and Higgs self-interactions and can therefore help distinguish a linear SMEFT description from the more general nonlinear HEFT framework. The phenomenology of these processes in HEFT, including higher-order corrections and comparisons with SMEFT, has been investigated in Refs.~\cite{Herrero:2022krh,Anisha:2024ljc,Anisha:2024ryj,Braun:2025hvr}.

The matching obtained in this work predicts a definite pattern of interactions relevant to these processes. At leading chiral order, the Higgs-dependent functions multiplying $\braket{V_\mu V^\mu}$ and $\braket{T V_\mu}\braket{T V^\mu}$ determine the interactions of two longitudinal gauge bosons with arbitrary powers of the Higgs field, while the matched potential $\mathcal{V}(h)$ determines the Higgs self-interactions. These structures contribute directly to double- and multi-Higgs production through vector-boson fusion. At $\mathcal{O}(p^4)$, the four-current operators, including $\braket{V_\mu V^\mu}\braket{V_\nu V^\nu}$ and $\braket{V_\mu V_\nu}\braket{V^\mu V^\nu}$ contribute to longitudinal VBS. Operators containing two Goldstone currents and two Higgs derivatives instead contribute to double-Higgs channels such as $W^+ W^-\to hh$ and $ZZ\to hh$ VBS and multi-Higgs measurements thus probe complementary sectors of the matched HEFT Lagrangian.

A characteristic feature of the triplet extension is the simultaneous generation of custodial-preserving and custodial-symmetry-breaking structures. The latter are identified by insertions of the spurion $T=U\sigma_3 U^\dagger$ and distinguish charged and neutral vector-boson interactions. A useful comparison is provided by the HEFT matching of the real Higgs-triplet extension in Ref.~\cite{Song:2025kjp}, where the resulting structures were studied through the processes $hh\to hh$, $W^+ W^-\to hh$, and $ZZ\to hh$. The complex triplet considered here contains, in addition, a CP-odd neutral scalar and a doubly charged scalar. It therefore generates a distinct and more extensive pattern of custodial-preserving and custodial-symmetry-breaking coefficients. Comparisons among the VBS channels $W^+ W^-\to W^+ W^-$, $W^+ W^-\to ZZ$, and $ZZ\to ZZ$ together with the double-Higgs channels $W^+ W^-\to hh$ and $ZZ\to hh$ can consequently probe both nonlinear Higgs dynamics and the specific pattern of custodial-symmetry breaking induced by the complex triplet. In particular, the combined analysis of charged and neutral channels can resolve correlations that would not be visible in a phenomenological treatment restricted to custodial symmetric Higgs interactions.

The same operator structures are closely related to the phenomenological description in terms of anomalous gauge couplings. After going to the unitary gauge, the four-current operators generate anomalous quartic gauge couplings. More generally, the complete HEFT basis also contains operators involving electroweak field strengths, which can modify anomalous triple gauge couplings and generate correlated quartic gauge interactions~\cite{Brivio:2016fzo,Eboli:2021unw,Eboli:2023mny}. HEFT therefore organizes anomalous gauge interactions and their associated Higgs couplings within a common symmetry-based framework, rather than treating them as unrelated broken-phase vertices. For the specific tree-level matching performed in this work, however, no independent bosonic operator containing an electroweak field strength is generated through $\mathcal O(p^4)$. At this order, the four-current operators therefore generate anomalous quartic gauge interactions, but no independent field-strength-induced anomalous triple gauge couplings. The latter may arise once loop-level matching is included. In principle, a combined analysis of VBS, multi-Higgs production, and anomalous quartic gauge couplings could constrain the correlations and custodial-symmetry-breaking pattern predicted by the type-II seesaw matching. Establishing whether the predicted deviations could be resolved experimentally requires a dedicated sensitivity study, which is not attempted here.

\subsection{Top-quark probes of Higgs and electroweak interactions}
Processes involving top quarks provide complementary probes of the nonlinear Higgs and electroweak interactions discussed above. Owing to the large top Yukawa coupling, the high-energy scattering processes $W_L^+ W_L^-\to t\bar{t}$, $Z_L Z_L\to t\bar{t}$ are particularly sensitive to departures from the SM relations among the Higgs--top, gauge--top, and Goldstone--top interactions. In the matched HEFT, these amplitudes receive contributions from the modified Yukawa function $\mathcal{Y}_Q(h)$ and from two-fermion operators containing Goldstone currents. The latter include both custodial-preserving structures and operators with insertions of $T$. Comparing the charged and neutral channels can therefore help disentangle the corresponding custodial-symmetry-breaking interactions. The high-energy amplitudes for vector-boson scattering into top-quark pairs have been studied within nonlinear electroweak EFT, while their collider sensitivity to anomalous top--Higgs and electroweak interactions has also been investigated in weak-boson-fusion production at high-energy muon colliders~\cite{Castillo:2016erh,Chen:2022yiu,Liu:2023yrb}.

Interactions involving two Higgs bosons probe a complementary part of the matched fermionic sector. In particular, the associated production process $pp\to t\bar{t}hh$ is sensitive to the nonderivative $t\bar{t}hh$ contact interaction generated by the expansion of $\mathcal{Y}_Q(h)$. It also probes the momentum-dependent $t\bar{t}hh$ interaction arising from two-fermion operators of the form $\bigl(\bar Q_L U Q_R\bigr)D_\mu hD^\mu h$ as well as contributions involving the $t\bar{t}h$ coupling and the Higgs self-interactions. The process therefore provides complementary sensitivity to the Yukawa function, derivative two-fermion operators, and the Higgs potential obtained from the matching. Associated $t\bar{t}hh$ production has recently been studied within HEFT as a probe of these interactions at hadron colliders~\cite{Matheus:2026dmj}.

Compared with purely bosonic VBS and multi-Higgs production, the correlated analysis of top-quark channels remains less developed within HEFT. Existing studies have mainly focused either on high-energy vector-boson-fusion amplitudes or on selected anomalous top--Higgs couplings. To our knowledge, a systematic collider analysis incorporating the correlated custodial-preserving and custodial-symmetry-breaking interactions predicted by the present top-down matching has not yet been performed. A combined study of weak-boson-fusion top-pair production and associated $t\bar{t}hh$ production would therefore provide complementary tests of the Yukawa and two-fermion sectors of the matched HEFT.

\section{Conclusion and Discussion\label{sec:con}}
In this work, we have constructed a complete and nonredundant basis of lepton-number-violating (LNV) HEFT operators through $\mathcal O(p^4)$, retaining their full flavor structures. Inspired by the Weinberg operator, we introduced the dressed spurion $\hat{\mathcal S}$ to incorporate the $B-L$ charge and custodial orientation of the LNV interactions. The resulting basis contains the leading Majorana-mass operator and higher-order two- and four-fermion interactions with $\Delta L=2$ and $\Delta L=4$. It also includes baryon-number-violating LNV operators, which, together with the previously known $B-L$-conserving $LQ^3$ class and their Hermitian conjugates, complete the classification for the light-field content and chiral orders considered here. The explicit operator reduction and flavor counting agree with the corresponding direct Hilbert-series enumeration~\cite{Graf:2022rco}. We further constructed a Hilbert series refined by $\hat{\mathcal S}$ and $\hat{\mathcal S}^{\dagger}$, which retains the $B-L$ grading absent from the conventional neutral $T$-spurion counting. For the minimal electrically neutral LNV orientation, the relation $\hat{\mathcal S}^{\dagger}\hat{\mathcal S}=(1+T)/2$ also makes explicit the connection between the LNV insertion and custodial-symmetry breaking, although the two spurion refinements cannot be identified term by term in the $B-L$-violating sectors.

As a concrete application, we have performed the complete tree-level matching of the type-II seesaw model onto HEFT through $\mathcal O(p^4)$, including the bosonic, lepton-number-conserving fermionic, and LNV sectors. Using the nonlinear field representation and the primary-HEFT power-counting scheme~\cite{Song:2024kos,Ge:2026qfa}, we integrated out the heavy states $K$, $\eta$, $H^\pm$, and $H^{\pm\pm}$ and reduced the resulting interactions by integration by parts and local field redefinitions. The matching retains the dependence on the independent heavy-scalar masses, the neutral-scalar mixing angle, and the triplet VEV without additional expansions in these parameters, and includes the complete flavor structures of the two- and four-fermion operators. While the operator structures are matched through the stated chiral order, their Higgs-dependent coefficient functions are displayed to the indicated field orders. Compared with the real-triplet extension~\cite{Song:2025kjp}, the additional CP-odd and doubly charged scalars lead to a different pattern of HEFT Wilson coefficients. We also recover the overlapping coefficients of the broken-phase EFT after translating the parameter conventions and expanding in the triplet-VEV ratio, while the original HEFT expressions retain their unexpanded parameter dependence.

The matched coefficients connect the type-II seesaw parameters to Higgs and electroweak precision observables, vector-boson scattering, multi-Higgs production, anomalous gauge interactions, and top-quark processes. In the lepton sector, we have discussed neutrinoless double-beta decay, charged-lepton flavor violation in $\mu\to3e$, and same-sign dilepton production. We have also provided the HEFT-to-LEFT matching relevant to neutrinoless double-beta decay, so that the resulting coefficients can be used in low-energy analyses and tools such as $\nu$DoBe~\cite{Scholer:2023bnn}. The type-II matching correlates the Majorana neutrino mass with higher-order LNV interactions. In particular, the nonresonant same-sign dilepton contact interaction generated by heavy $H^{\pm\pm}$ exchange is controlled by the triplet Yukawa coupling and VEV combination entering the neutrino mass matrix, and is consequently strongly suppressed in the minimal heavy-triplet scenario. Near the heavy-scalar resonances, the full model rather than its local HEFT description is required.

Our analysis is restricted to tree-level matching through $\mathcal O(p^4)$. A next-to-leading-order treatment would require one-loop matching and renormalization, together with the evolution of Wilson coefficients from the heavy-scalar scale to the electroweak and low-energy scales. It would also generate operators absent at tree level, such as dipole interactions relevant for radiative charged-lepton-flavor-violating decays. Extending the LNV basis and matching to higher chiral orders would permit a more complete analysis of short-range contributions to neutrinoless double-beta decay. Finally, applying the same matching procedure to other scalar extensions would allow the patterns of HEFT Wilson coefficients associated with different electroweak representations to be compared. The general LNV operator basis and the type-II seesaw matching presented here provide the ingredients for such studies within a common nonlinear electroweak framework.

\acknowledgments
H.S. would like to thank Hao Sun for useful discussions on the construction of the LNV operator basis in HEFT. The work of H.S. is supported by IBS under the project code, IBS-R018-D1. The work of Z.G. and X.W. is supported by
the National Science Foundation of China under Grants No. 11947416.

\appendix
\section{A proof of the redundancy of $\mathcal{O}^{R; \bar{L}R}_{B1}+\mathcal{O}^{R; \bar{L}R}_{B2}$~\label{app:TypeBred}}
We first note that the spurions $T$ and $S^\dagger$ satisfy the following identity,
\begin{align}
    (\epsilon S^\dagger\epsilon)_{ab}{(1+T)^c}_d=-\frac{1}{2}\epsilon_{da}{(S^\dagger\epsilon)^c}_b+\frac{1}{2}(\epsilon T)_{da}{(S^\dagger\epsilon)^c}_b-\frac{1}{2}\epsilon_{db}{(S^\dagger\epsilon)^c}_a+\frac{1}{2}(\epsilon T)_{db}{(S^\dagger\epsilon)^c}_a
\end{align}
The left-hand side is exactly the Type-B weak tensor
\begin{align}
    (\epsilon S^\dagger\epsilon)\otimes(1+T),
\end{align}
while the right-hand side is a sum of Type-A weak tensors
\begin{align}
    \epsilon\otimes(S^\dagger\epsilon),\qquad\epsilon T\otimes(S^\dagger\epsilon).
\end{align}
Substituting the identity into $\mathcal{O}^{R; \bar{L}R}_{B1, prst}+\mathcal{O}^{R; \bar{L}R}_{B2, prst}$, we obtain
\begin{align}
    \mathcal{O}^{R; \bar{L}R}_{B1, prst}+\mathcal{O}^{R; \bar{L}R}_{B2, prst}&=\epsilon_{\alpha\beta\gamma}[(U Q_{R p}^\alpha)^{a T}C(U Q_{R r}^\beta)^b][\bar{L}_{L s c}(U Q_{R t}^\gamma)^d](\epsilon S^\dagger\epsilon)_{ab}{(1+T)^c}_d \nonumber \\
    &=X_1+X_2+X_3+X_4
\end{align}
with
\begin{align}
    X_1=&-\frac{1}{2}\epsilon_{\alpha\beta\gamma}[(U Q_{R p}^\alpha)^{a T}C(U Q_{R r}^\beta)^b][\bar{L}_{L s c}(U Q_{R t}^\gamma)^d]\epsilon_{da}{(S^\dagger\epsilon)^c}_b \nonumber \\
    X_2=&+\frac{1}{2}\epsilon_{\alpha\beta\gamma}[(U Q_{R p}^\alpha)^{a T}C(U Q_{R r}^\beta)^b][\bar{L}_{L s c}(U Q_{R t}^\gamma)^d](\epsilon T)_{da}{(S^\dagger\epsilon)^c}_b \nonumber \\
    X_3=&-\frac{1}{2}\epsilon_{\alpha\beta\gamma}[(U Q_{R p}^\alpha)^{a T}C(U Q_{R r}^\beta)^b][\bar{L}_{L s c}(U Q_{R t}^\gamma)^d]\epsilon_{db}{(S^\dagger\epsilon)^c}_a \nonumber \\
    X_4=&+\frac{1}{2}\epsilon_{\alpha\beta\gamma}[(U Q_{R p}^\alpha)^{a T}C(U Q_{R r}^\beta)^b][\bar{L}_{L s c}(U Q_{R t}^\gamma)^d](\epsilon T)_{db}{(S^\dagger\epsilon)^c}_a
\end{align}
After applying the same-chirality three-quark Fierz/color identity
\begin{align}
    \epsilon_{\alpha\beta\gamma}q_{p,\Dot{\alpha}}^{\alpha a}q_{r,\Dot{\beta}}^{\beta b}q_{t,\Dot{\gamma}}^{\gamma d}\bar{\ell}_{sc\Dot{\delta}}\times(\epsilon^{\Dot{\alpha}\Dot{\beta}}\epsilon^{\Dot{\gamma}\Dot{\delta}}+\epsilon^{\Dot{\beta}\Dot{\gamma}}\epsilon^{\Dot{\alpha}\Dot{\delta}}+\epsilon^{\Dot{\gamma}\Dot{\alpha}}\epsilon^{\Dot{\beta}\Dot{\delta}})=0
\end{align}
where $q_{p,\Dot{\alpha}}^{\alpha a},\,q_{r,\Dot{\beta}}^{\beta b},\,q_{t,\Dot{\gamma}}^{\gamma d}$ are three right-handed quark fields, with color indices $\alpha,\,\beta,\,\gamma$, weak indices $a,\, b,\, c$, and dotted spinor indices $\Dot{\alpha},\,\Dot{\beta},\,\Dot{\gamma}$ and $\bar{\ell}_{sc\Dot{\delta}}$ is a left-handed lepton, or equivalently, in bilinear notation
\begin{align}
    \epsilon_{\alpha\beta\gamma}(q_p^{\alpha a T}C q_r^{\beta b})(\bar{L}_{L s c}q_t^{\gamma d})+\epsilon_{\alpha\beta\gamma}(q_r^{\alpha b T}C q_t^{\beta d})(\bar{L}_{L s c}q_p^{\gamma a})+\epsilon_{\alpha\beta\gamma}(q_t^{\alpha d T}C q_p^{\beta a})(\bar{L}_{L s c}q_r^{\gamma b})=0
\end{align}
we find
\begin{align}
    X_1=&+\frac{1}{2}\epsilon_{\alpha\beta\gamma}[(U Q_{R t}^\alpha)^{d T}C(U Q_{R p}^\beta)^a][\bar{L}_{L s c}(U Q_{R t}^\gamma)^b]\epsilon_{da}{(S^\dagger\epsilon)^c}_b \nonumber \\
    &+\frac{1}{2}\epsilon_{\alpha\beta\gamma}[(U Q_{R r}^\alpha)^{b T}C(U Q_{R t}^\beta)^d][\bar{L}_{L s c}(U Q_{R p}^\gamma)^a]\epsilon_{da}{(S^\dagger\epsilon)^c}_b=\frac{1}{2}\mathcal{O}^{R; \bar{L}R}_{A1, tpsr}+X_{1b} \\
    X_2=&-\frac{1}{2}\epsilon_{\alpha\beta\gamma}[(U Q_{R t}^\alpha)^{d T}C(U Q_{R p}^\beta)^a][\bar{L}_{L s c}(U Q_{R r}^\gamma)^b](\epsilon T)_{da}{(S^\dagger\epsilon)^c}_b \nonumber \\
    &-\frac{1}{2}\epsilon_{\alpha\beta\gamma}[(U Q_{R r}^\alpha)^{b T}C(U Q_{R t}^\beta)^d][\bar{L}_{L s c}(U Q_{R p}^\gamma)^a](\epsilon T)_{da}{(S^\dagger\epsilon)^c}_b=-\frac{1}{2}\mathcal{O}^{R; \bar{L}R}_{A2, tpsr}+X_{2b} \\
    X_3=&+\frac{1}{2}\epsilon_{\alpha\beta\gamma}[(U Q_{R t}^\alpha)^{d T}C(U Q_{R p}^\beta)^a][\bar{L}_{L s c}(U Q_{R r}^\gamma)^b]\epsilon_{db}{(S^\dagger\epsilon)^c}_a \nonumber \\
    &+\frac{1}{2}\epsilon_{\alpha\beta\gamma}[(U Q_{R r}^\alpha)^{b T}C(U Q_{R t}^\beta)^d][\bar{L}_{L s c}(U Q_{R p}^\gamma)^a]\epsilon_{db}{(S^\dagger\epsilon)^c}_a=X_{3a}+\frac{1}{2}\mathcal{O}^{R; \bar{L}R}_{A1, rtsp} \\
    X_4=&-\frac{1}{2}\epsilon_{\alpha\beta\gamma}[(U Q_{R t}^\alpha)^{d T}C(U Q_{R p}^\beta)^a][\bar{L}_{L s c}(U Q_{R r}^\gamma)^b](\epsilon T)_{db}{(S^\dagger\epsilon)^c}_a \nonumber \\
    &-\frac{1}{2}\epsilon_{\alpha\beta\gamma}[(U Q_{R r}^\alpha)^{b T}C(U Q_{R t}^\beta)^d][\bar{L}_{L s c}(U Q_{R p}^\gamma)^a](\epsilon T)_{db}{(S^\dagger\epsilon)^c}_a=X_{4a}-\frac{1}{2}\mathcal{O}^{R; \bar{L}R}_{A2, rtsp}
\end{align}
where $X_{1b},\, X_{2b},\, X_{3a}$ and $X_{4a}$ are not immediately of Type-A form since since their weak-index contractions are crossed relative to the Lorentz bilinears. However, we can further recombine the crossed pieces
\begin{align}
    X_{1b}+X_{2b}&=\frac{1}{2}\epsilon_{\alpha\beta\gamma}[(U Q_{R r}^\alpha)^{b T}C(U Q_{R t}^\beta)^d][\bar{L}_{L s c}(U Q_{R p}^\gamma)^a]\times[\epsilon_{da}-(\epsilon T)_{da}]{(S^\dagger\epsilon)^c}_b \\
    X_{3a}+X_{4a}&=\frac{1}{2}\epsilon_{\alpha\beta\gamma}[(U Q_{R t}^\alpha)^{d T}C(U Q_{R p}^\beta)^a][\bar{L}_{L s c}(U Q_{R r}^\gamma)^b]\times[\epsilon_{db}-(\epsilon T)_{db}]{(S^\dagger\epsilon)^c}_a
\end{align}
Using the weak identities
\begin{align}
    [\epsilon_{da}-(\epsilon T)_{da}]{(S^\dagger\epsilon)^c}_b=&-[\epsilon_{bd}+(\epsilon T)_{bd}]{(S^\dagger\epsilon)^c}_a \\
    [\epsilon_{db}-(\epsilon T)_{db}]{(S^\dagger\epsilon)^c}_a=&+[\epsilon_{da}-(\epsilon T)_{da}]{(S^\dagger\epsilon)^c}_b
\end{align}
we get
\begin{align}
    X_{1b}+X_{2b}=&-\frac{1}{2}\mathcal{O}^{R; \bar{L}R}_{A1, rtsp}-\frac{1}{2}\mathcal{O}^{R; \bar{L}R}_{A2, rtsp} \\
    X_{3a}+X_{4a}=&+\frac{1}{2}\mathcal{O}^{R; \bar{L}R}_{A1, tpsr}-\frac{1}{2}\mathcal{O}^{R; \bar{L}R}_{A2, tpsr}
\end{align}
Adding everything, we find
\begin{align}
    \mathcal{O}^{R; \bar{L}R}_{B1, prst}+\mathcal{O}^{R; \bar{L}R}_{B2, prst}=\mathcal{O}^{R; \bar{L}R}_{A1, tpsr}-\mathcal{O}^{R; \bar{L}R}_{A2, tpsr}-\mathcal{O}^{R; \bar{L}R}_{A2, rtsp}
\end{align}
which indicates one linear combination of $\mathcal{O}^{R; \bar{L}R}_{B1}$ and $\mathcal{O}^{R; \bar{L}R}_{B2}$ is in the space spanned by Type-A operators.

\section{HEFT-to-LEFT matching for neutrinoless double-beta decay~\label{app:0vbb_matching}}
In this appendix, we provide the explicit matching between the $\Delta L=2$ HEFT operators relevant to neutrinoless double-beta decay and the corresponding LEFT operator basis~\cite{Jenkins:2017jig}. After electroweak symmetry breaking, we work in the unitary gauge and evaluate the HEFT coefficient functions at $h=0$, since no external Higgs field is involved in the decay. For each HEFT operator, the tables list the associated LEFT operator—or operators—the corresponding Wilson-coefficient relation, and the code label adopted by $\nu$DoBe~\cite{Scholer:2023bnn}. The resulting LEFT coefficients can therefore be used directly as electroweak-scale input for calculations of neutrinoless double-beta-decay observables within $\nu$DoBe, after selecting the appropriate electron and first-generation quark flavor indices. At the chiral orders considered here, the matching includes the Majorana-mass contribution and the associated lower-order long-range interactions, whereas a complete treatment of the short-range six-fermion sector requires extending the HEFT basis to higher chiral orders.
\begin{table}[ht!]
    \centering
    \resizebox{\textwidth}{!}{%
    \begin{tabular}{|l|l|r|r|r|}
        \hline\textbf{Class} & \multicolumn{1}{c|}{\textbf{Op. Name}} & \multicolumn{1}{c|}{\textbf{Op. Structure}} & \multicolumn{1}{c|}{\textbf{LEFT Op.}} & \multicolumn{1}{c|}{\textbf{Matching Condition}} \\ \hline
        
        \multicolumn{5}{c}{$\mathcal{O}(p^2)$} \\ \hline
        
        $\psi^2$ & $\mathcal{O}_{L}$ & $L_{L p}^T C\hat{\mathcal{S}}L_{L r}$ & $m_{\beta\beta}$ & $m_{\beta\beta}^{pr}=-\mathcal{F}_{L}^{pr}(0)$\\ \hline
        
        \multicolumn{5}{c}{$\mathcal{O}(p^3)$} \\ \hline
        
        $\psi^2 V$ & $\mathcal{O}_{LV}$ & $L_{L p}^T C\hat{\mathcal{S}}\gamma^\mu V_\mu U L_{R r}$ & $O_{VL}^{(6)}$ & $C_{VL}^{(6)\, prst}\supset -2iV_{st}\mathcal{F}_{LV}^{rp*}(0)$ \\ \hline
        
        \multicolumn{5}{c}{$\mathcal{O}(p^4)$} \\ \hline
        
        $\psi^2 W$ & $\mathcal{O}_{LX,W}$ & $L_{L p}^T C\hat{\mathcal{S}}\sigma^{\mu\nu}W_{\mu\nu}L_{L r}$ & $O_{VL}^{(6)},\, O_{VL}^{(7)},\, O_{VL}^{(9)}$ & \makecell[r]{$C_{VL}^{(6)\, prst}\supset -\frac{4}{g}V_{st}(M_e^\dagger)_{pu}\mathcal{F}_{LX,W}^{ru*}(0)$ \\ $C_{VL}^{(7)\, prst}\supset -\frac{4v}{g}V_{st}\mathcal{F}_{LX,W}^{rp*}(0)$ \\ $C_{1L}^{(9)\, prstuv}\supset -\frac{8v}{g}V_{st}V_{uv}\left[\mathcal{F}_{LX,W}^{rp*}(0)\right.$ \\ $\left.+\mathcal{F}_{LX,W}^{pr*}(0)\right]$} \\ \hline
        
        \multirow{2}{*}{$\psi^2 V^2$} & $\mathcal{O}_{LVV,4}$ & $L_{L p}^T C\epsilon\hat{\mathcal{S}}^\dagger\epsilon L_{L r}\braket{\epsilon\hat{\mathcal{S}}V_\mu}\braket{\epsilon\hat{\mathcal{S}}V^\mu}$ & $O_{1L}^{(9)}$ & $C_{1L}^{(9)\, prstuv}\supset 4v V_{st}V_{uv}\mathcal{F}_{LVV,4}^{pr*}(0)$ \\
         & $\mathcal{O}_{LVV,7}$ & $L_{R p}^T C U^T\epsilon\hat{\mathcal{S}}^\dagger\epsilon U L_{R r}\braket{\epsilon\hat{\mathcal{S}}V_\mu}\braket{\epsilon\hat{\mathcal{S}}V^\mu}$ & $O_{1R}^{(9)}$ & $C_{1R}^{(9)\, prstuv}\supset 4v V_{st}V_{uv}\mathcal{F}_{LVV,7}^{pr*}(0)$ \\ \hline
         
        \multirow{8}{*}{$\psi^4$} & $\mathcal{O}_{SQ,LR}^{\epsilon}$ & $(L_{L p}^T C\epsilon L_{L r})(\bar{Q}_{L s}\epsilon\hat{\mathcal{S}}U Q_{R t})$ & $O_{SL}^{(6)}$ & $C_{SL}^{(6)\, prst}\supset v^2\mathcal{F}_{SQ,LR}^{\epsilon\, prts*}(0)\rightarrow 0$ \\
         & $\mathcal{O}_{SQ,LR}^{\epsilon T}$ & $(L_{L p}^T C\epsilon T L_{L r})(\bar{Q}_{L s}\epsilon\hat{\mathcal{S}}U Q_{R t})$ & $O_{SL}^{(6)}$ & $C_{SL}^{(6)\, prst}\supset v^2\mathcal{F}_{SQ,LR}^{\epsilon T\, prts*}(0)$ \\
         
         & $\mathcal{O}_{SQ,RL}^{\epsilon}$ & $(L_{L p}^T C\epsilon L_{L r})(\bar{Q}_{R s}U^\dagger \epsilon\hat{\mathcal{S}}Q_{L t})$ & $O_{SR}^{(6)}$ & $C_{SR}^{(6)\, prst}\supset v^2\mathcal{F}_{SQ,RL}^{\epsilon\, prts*}(0)\rightarrow 0$ \\
         & $\mathcal{O}_{SQ,RL}^{\epsilon T}$ & $(L_{L p}^T C\epsilon T L_{L r})(\bar{Q}_{R s}U^\dagger\epsilon\hat{\mathcal{S}}Q_{L t})$ & $O_{SR}^{(6)}$ & $C_{SR}^{(6)\, prst}\supset v^2\mathcal{F}_{SQ,RL}^{\epsilon T\, prts*}(0)$ \\
         
         & $\mathcal{O}_{TQ,RL}^{\epsilon}$ & $(L_{L p}^T C\sigma^{\mu\nu}\epsilon L_{L r})(\bar{Q}_{R s}\sigma_{\mu\nu}U^\dagger \epsilon\hat{\mathcal{S}}Q_{L t})$ & $O_{T}^{(6)}$ & $C_{T}^{(6)\, prst}\supset v^2\mathcal{F}_{TQ,RL}^{\epsilon\, prts*}(0)\rightarrow 0$ \\
         & $\mathcal{O}_{TQ,RL}^{\epsilon T}$ & $(L_{L p}^T C\sigma^{\mu\nu}\epsilon T L_{L r})(\bar{Q}_{R s}\sigma_{\mu\nu}U^\dagger\epsilon\hat{\mathcal{S}}Q_{L t})$ & $O_{T}^{(6)}$ & $C_{T}^{(6)\, prst}\supset v^2\mathcal{F}_{TQ,RL}^{\epsilon T\, prts*}(0)$ \\
         
         & $\mathcal{O}_{VQ,L}$ & $(L_{L p}^T C\gamma^\mu\epsilon U L_{R r})(\bar{Q}_{L s}\gamma_\mu\epsilon\hat{\mathcal{S}}Q_{L t})$ & $O_{VL}^{(6)}$ & $C_{VL}^{(6)\, prst}\supset -v^2\mathcal{F}_{VQ,L}^{rpts*}(0)$ \\
         & $\mathcal{O}_{VQ,R}$ & $(L_{L p}^T C\gamma^\mu\epsilon U L_{R r})(\bar{Q}_{R s}U^\dagger\gamma_\mu\epsilon\hat{\mathcal{S}}U Q_{R t})$ & $O_{VR}^{(6)}$ & $C_{VR}^{(6)\, prst}\supset -v^2\mathcal{F}_{VQ,R}^{rpts*}(0)$ \\ \hline
    \end{tabular}%
    }
    \caption{Matching of the $\Delta L=2$ HEFT operators relevant to neutrinoless double-beta decay onto the Majorana-mass parameter and the corresponding LEFT operators. The LEFT operators are written in the notation adopted by $\nu$DoBe~\cite{Scholer:2023bnn}. The final column gives the matching relations between the LEFT Wilson coefficients and the HEFT coefficient functions evaluated at $h=0$. Flavor indices are displayed explicitly. The notation $\to 0$ indicates that the corresponding contribution vanishes after specializing to the flavor assignment relevant for $0\nu\beta\beta$, in particular $p=r=e$, owing to the antisymmetry of the associated HEFT operator under $p\leftrightarrow r$; its general-flavor matching does not vanish.~\label{tab:HEFT_LEFT_0nubb}}
\end{table}

\section{Detailed comparison with the broken-phase effective theory\label{app:bEFT_details}}
This appendix contains the coefficient-level comparison referred to in Sec.~\ref{sec:bEFT_comparison}. The second column gives the unexpanded HEFT results in our conventions and the third gives the corresponding coefficients of Ref.~\cite{Liao:2025bmn}; comparisons are made only after the respective parameter conventions are identified and both expressions are expanded consistently in $\epsilon=v_\Xi/v_{\rm EW}$. The presence of heavy masses and mixing parameters in the exact HEFT expressions should not be confused with the order to which the comparison is performed.

\newcommand{\vew}{v_{\rm ew}}
\newcommand{\vH}{\vh}
\newcommand{\cw}{c_W}
\newcommand{\eps}{\epsilon}
\newcommand{\rd}{r_\Delta}
\newcommand{\x}{\xi}
\newcommand{\ca}{c}
\newcommand{\sa}{s}
\newcommand{\lm}{\lambda_-}
\newcommand{\mK}{m_K}
\newcommand{\meta}{m_\eta}
\newcommand{\mpm}{m_+}
\newcommand{\mpp}{m_{++}}

%
%
%
%
%
%
%
%

\begin{table*}[htbp]
\centering
\caption{Comparison of the broken-phase interactions of $\textbf{dim-3}$ obtained from our HEFT matching with the corresponding bEFT results in Table~1 of Ref.~\cite{Liao:2025bmn}. We use $\mathcal{N}_{pr}(h)$ to denote the Wilson coefficient function of Weinberg operator $\mathcal{O}_L^{pr}=L_{Lp}^T C\hat{\mathcal{S}}L_{Lr}$. Here and throughout, the superscript $(n)$ denotes the $n$th derivative of the corresponding function with respect to $h/v_{\rm EW}$, evaluated at the argument indicated. The first column gives the broken-phase interaction, while the second column presents the relations between the corresponding coefficients and the HEFT Wilson coefficient functions, together with their exact expressions in our notation prior to expanding in $\epsilon\equiv v_\Xi/v_{\rm EW}$. The third column shows the bEFT coefficient in the conventions of Ref.~\cite{Liao:2025bmn}. Agreement is obtained after translating the parameter conventions and expanding the HEFT result consistently in $\epsilon$. We also define $g_Z\equiv g/c_W$ for simplicity.}
\label{tab:bEFT_comparison}
\begin{tabular}{ccc}
\toprule
\multicolumn{3}{c}{$\textbf{dim-3}$} \\
\midrule
\makecell[c]{Operator relation} &
\makecell[c]{Exact WC in our notation\footnote{Here $Y^{pr}\equiv\sqrt{2}\,y^{pr}$, introduced for comparison.}} &
\makecell[c]{bEFT WC of Ref.~\cite{Liao:2025bmn}} \\
\midrule
\multicolumn{3}{c}{$\Delta L=0$} \\
\midrule
\makecell[c]{$\displaystyle h^3$} &
\makecell[c]{$-\frac{1}{3!}\mathcal{V}^{(3)}(0)$ \\
$=-\frac{\sa^2\meta^2(2\x\ca+\sa)}{2\x(4\x^2+1)\vh}+\frac{m_h^2(\sa^3-\x\ca^3)}{2\x\vh}$} & \makecell[c]{$-\frac{\lambda}{2}\vew+\eps^2\vew\Bigl[\frac{(1-2\kappa^2)\lm}{2}-\frac{1-\kappa+2\kappa^2-6\kappa^3}{2\kappa\rd^2}\Bigr]$}
\\
\makecell[c]{$\displaystyle hW^-_\mu W^{+\mu}$} &
\makecell[c]{$\frac{1}{4}g^2\vev^2\mathcal{F}^{(1)}(0)$ \\
$=\frac{g^2}{2}\vh(\ca-2\x\sa)$} &
\makecell[c]{$\displaystyle \frac{g^2\vew}{2}\displaystyle \left[1-\eps^2(1-4\kappa+2\kappa^2)\right]$}
\\
\makecell[c]{$\displaystyle hZ_\mu Z^\mu$} &
\makecell[c]{$\frac{1}{8}g_Z^2\vev^2\mathcal{F}^{(1)}(0)+\frac{1}{4}g_Z^2\vev^2\mathcal{G}^{(1)}(0)$ \\
$=\frac{g^2}{4\cw^2}\vh(\ca-4\x\sa)$} &
\makecell[c]{$\displaystyle \frac{g^2\vew}{4\cw^2}\displaystyle \left[1-\eps^2(1-8\kappa+2\kappa^2)\right]$}
\\
\midrule
\multicolumn{3}{c}{$\Delta L=2$} \\
\midrule
\makecell[c]{$\displaystyle \overline{\nu^C_{L,p}}\nu_{L,r}$} & \makecell[c]{$\mathcal{N}_{pr}(0)=\displaystyle -\frac{\x\vh}{2}Y_\Delta^{pr}$} & \makecell[c]{$\displaystyle -\frac{\eps\vew}{2}Y_\Delta^{pr}$} \\
\bottomrule
\end{tabular}
\end{table*}

\begin{table*}[htbp]
\centering
\caption{Similar to TABLE~\ref{tab:bEFT_comparison}. Comparison of the broken-phase interactions of $\textbf{dim-4}$ obtained from our HEFT matching with the corresponding bEFT results in Table~2 of Ref.~\cite{Liao:2025bmn}. For simplicity, we use $\mathcal{F}_{1,2,3,4,5}(h)$ to denote the Wilson coefficient functions associated with the bosonic operators $\braket{V_\mu V^\mu}\braket{V_\nu V^\nu}$, $\braket{V_\mu V_\nu}\braket{V^\mu V^\nu}$, $\braket{T V_\mu}\braket{T V^\mu}\braket{V_\nu V^\nu}$, $\braket{T V_\mu}\braket{T V_\nu}\braket{V^\mu V^\nu}$, and $\braket{T V_\mu}\braket{T V^\mu}\braket{T V_\nu}\braket{T V^\nu}$, respectively. We also define $g_Z\equiv g/c_W$ for simplicity.}
\label{tab2:bEFT_comparison}
\resizebox{\linewidth}{!}{
\begin{tabular}{ccc}
\toprule
\multicolumn{3}{c}{$\textbf{Dim. 4}$} \\
\midrule
\makecell[c]{Operator relation} &
\makecell[c]{Exact WC in our notation} &
\makecell[c]{bEFT WC of Ref.~\cite{Liao:2025bmn}} \\
\midrule
\multicolumn{3}{c}{$\Delta L=0$} \\
\midrule

\makecell[l]{$\displaystyle h^4$} &
\makecell[c]{$-\frac{1}{4!}\mathcal{V}^{(4)}(0)$ \\ $=\frac{s^2 m_\eta^2}{8 \xi^2(4 \xi^2 + 1)\vh^2}\biggl\{\frac{m_\eta^2\bigl[3 s(c - 2 \xi s) + 4 \xi\bigr]^2}{(4 \xi^2 + 1)m_K^2}+ 2 \xi c(9 s^2 - 7)s$ \\ $ + 6 s^4- 12 \xi^2(s^2 - 1)^2 - 5 s^2\biggr\} + \mathcal{O}(t^0)$} &
\makecell[c]{$\displaystyle -\frac{\lambda}{8}-\eps^2\left[
\frac{(1-\kappa)\kappa}{\rd^2}
+\kappa^2\lm \right.$ \\
$\displaystyle \left.
\qquad+\frac{(2-3\kappa+\kappa\lm\rd^2)^2}
{2(1+\lm\rd^2)\rd^2}
\right]$} \\

\makecell[l]{$\displaystyle h^2W^-_\mu W^{+\mu}$} &
\makecell[c]{$\frac{1}{4}g^2\vev^2\cdot\frac{1}{2!}\mathcal{F}^{(2)}(0)$ \\ $=\frac{g^2}{4}\biggl\{1 + s^2 - \frac{s(2 \xi c + s)}{\xi(4 \xi^2 + 1)m_K^2}\Bigl[c(4 \xi^2 + 1)m_K^2(\xi c + s)$ \\
$- m_\eta^2(3 c s + 4 \xi - 6 \xi s^2)\Bigr]\biggr\} + \mathcal{O}(t^1)$} &
\makecell[c]{$\displaystyle
\frac{g^2}{4}\left[
1-\eps^2
\frac{4(1-2\kappa)(2-\kappa+\kappa\lm\rd^2)}
{1+\lm\rd^2}
\right]$} \\

\makecell[l]{$\displaystyle h^2Z_\mu Z^\mu$} &
\makecell[c]{$\frac{1}{8}g_Z^2\vev^2\cdot\frac{1}{2!}\mathcal{F}^{(2)}(0)+\frac{1}{4}g_Z^2\vev^2\cdot\frac{1}{2!}\mathcal{G}^{(2)}(0)$ \\ $=\frac{g^2}{8\cw^2}\biggl\{1-2 s^2+5 s^4-4 c^3 \xi s-\frac{c s^3}{\xi}$ \\
$+ \frac{s m_\eta^2\bigl[8 c \xi^2(2 - 3 s^2) + 3 c s^2 + 2 \xi(8 - 9 s^2)s\bigr]}{\xi(4 \xi^2 + 1)m_K^2}\biggr\} + \mathcal{O}(t^1)$} &
\makecell[c]{$\displaystyle
\frac{g^2}{8\cw^2}
\left[
1-\eps^2
\frac{8(1-2\kappa)(2+\kappa\lm\rd^2)}
{1+\lm\rd^2}
\right]$} \\

\makecell[l]{$\displaystyle
W^-_\mu W^{+\mu}W^-_\nu W^{+\nu}$} &
\makecell[c]{$g^4\mathcal{F}_1(0)+\frac12 g^4\mathcal{F}_2(0)$ \\
$=\displaystyle
\frac{g^4}8\frac{\vh^2(2 \xi c + s)^2}{m_K^2} + \mathcal{O}(t^2)$} &
\makecell[c]{$\displaystyle
-\frac{g^4}{2}\left[
1-\eps^2
\frac{(1-\kappa)^2\rd^2}
{1+\lm\rd^2}
\right]$} \\

\makecell[l]{$\displaystyle
W^-_\mu W^+_\nu W^{-\mu}W^{+\nu}$} &
\makecell[c]{$\frac12 g^4\mathcal{F}_2(0)$ \\
$=\displaystyle
\frac{g^4}2\frac{\x^2\vh^2}{\mpp^2} + \mathcal{O}(t^2)$} &
\makecell[c]{$\displaystyle
\frac{g^4}{2}\left[
1+\eps^2
\frac{\rd^2}
{1+\lm\rd^2}
\right]$} \\

\makecell[l]{$\displaystyle
W^-_\mu W^{+\mu}Z_\nu Z^\nu$} &
\makecell[c]{$g^2g_Z^2\mathcal{F}_1(0)
+g^2g_Z^2\mathcal{F}_3(0)$ \\
$=\displaystyle
\frac{g^4}{8\cw^2}\frac{\vh^2(2\x\ca+\sa)(4\x\ca+\sa)}{\mK^2} + \mathcal{O}(t^2)$} &
\makecell[c]{$\displaystyle
-g^4\cw^2\left[
1-\eps^2
\frac{(2-\kappa)(1-\kappa)\rd^2}
{2\cw^4(1+\lm\rd^2)}
\right]$} \\

\makecell[l]{$\displaystyle
W^-_\mu W^+_\nu Z^\mu Z^\nu$} &
\makecell[c]{$g^2g_Z^2\mathcal{F}_2(0)
+g^2g_Z^2\mathcal{F}_4(0)$ \\
$=\displaystyle
\frac{g^4}{2\cw^2}
\frac{\x^2\vh^2}{(2\x^2+1)\mpm^2} + \mathcal{O}(t^2)$} &
\makecell[c]{$\displaystyle
g^4\cw^2\left[
1+\eps^2
\frac{\rd^2}
{\cw^4\left(2+(2\lm+\lambda_5)\rd^2\right)}
\right]$} \\

\makecell[l]{$\displaystyle
Z_\mu Z^\mu Z_\nu Z^\nu$} &
\makecell[c]{$\frac14 g_Z^4\mathcal{F}_1(0)
+\frac14 g_Z^4\mathcal{F}_2(0)+\frac12 g_Z^4\mathcal{F}_3(0)
+\frac12 g_Z^4\mathcal{F}_4(0)
+g_Z^4\mathcal{F}_5(0)$ \\ $=\displaystyle
\frac{g^4}{32\cw^4}\frac{\vh^2(4\x\ca+\sa)^2}{\mK^2} + \mathcal{O}(t^2)$} &
\makecell[c]{$\displaystyle
\eps^2
\frac{g^4(2-\kappa)^2\rd^2}
{8\cw^4(1+\lm\rd^2)}$} \\

\makecell[l]{$\displaystyle
h\bar\psi_{L,p}\psi_{R,r}$ \\
$\displaystyle (\psi=u,d,e)$} &
\makecell[c]{$\mathcal{Y}_{L,pr}^{(1)}(0)=\displaystyle
-\frac{\ca}{\sqrt2}\,Y_\psi^{pr}$} &
\makecell[c]{$\displaystyle
-\frac{1}{\sqrt2}\,Y_\psi^{pr}$} \\

\midrule
\multicolumn{3}{c}{$\Delta L=2$} \\
\midrule

\makecell[l]{$\displaystyle
h\left(\overline{\nu^C_{L,p}}\nu_{L,r}\right)$} &
\makecell[c]{$\mathcal{N}_{pr}^{(1)}(0)=\displaystyle
-\frac{\sa}{2}\,Y_\Delta^{pr}$} &
\makecell[c]{$\displaystyle
-\eps\,Y_\Delta^{pr}
(1+\lm\rd^2)$} \\
\bottomrule
\end{tabular}}
\end{table*}

\clearpage

\bibliographystyle{JHEP}
\bibliography{reference}

\end{document}